\documentclass[a4paper]{article}
\usepackage{amsmath,amssymb}
\usepackage{euscript}
\newtheorem{prp}{Proposition}[section]
\newtheorem{lem}[prp]{Lemma}
\newtheorem{thm}[prp]{Theorem}
\newtheorem{cor}[prp]{Corollary}

\newenvironment{prf}{\begin{trivlist} \item[{\em Proof.}]}%
 {\end{trivlist} \bigskip \par}
\newenvironment{sprf}{\begin{trivlist} \item[{\em Sketch of proof.}]}%
 {\end{trivlist} \bigskip \par}
\newenvironment{prfof}[1]{\begin{trivlist}\item[{{\em Proof of #1.}}]}%
 {\end{trivlist}\bigskip\par}
\newenvironment{rem}{\begin{trivlist} \item[{\em Remark.}]}%
  {\ignorespaces\end{trivlist} \bigskip \par}
\catcode`@=11
\def\qed{\relax\ifmmode\let\@tempa\relax\ifcase\@eqcnt\def\@tempa{& & &}\or%
\def\@tempa{& &}\else\def\@tempa{&}\fi\@tempa $\square$ \else\hfill%
{\footnotesize$\square$} \fi}
\@addtoreset{equation}{section}
\@addtoreset{figure}{section}
\catcode`@=12

\def\prpb{\begin{prp}}            \def\prpe{\end{prp}}
\def\lemb{\begin{lem}}            \def\leme{\end{lem}}
\def\thmb{\begin{thm}}            \def\thme{\end{thm}}
\def\corb{\begin{cor}}            \def\core{\end{cor}}
\def\excb{\begin{exc}}            \def\exce{\end{exc}}
\def\prfb{\begin{prf}}            \def\prfe{\end{prf}}
\def\sprfb{\begin{sprf}}          \def\sprfe{\end{sprf}}
\def\prfofb#1{\begin{prfof}{#1}}  \def\prfofe{\end{prfof}}
\def\remb{\begin{rem}}            \def\reme{\end{rem}}
\def\Itmb{\begin{Itemize}}        \def\Itme{\end{Itemize}}
\def\itmb{\begin{itemize}}        \def\itme{\end{itemize}}
\def\arb{\begin{eqnarray}}        \def\are{\end{eqnarray}}
\def\arsb{\begin{eqnarray*}}      \def\arse{\end{eqnarray*}}
\def\eqb#1\eqe{\begin{equation}#1\end{equation}}  
\def\eqsb#1\eqse{\begin{equation*}#1\end{equation*}}  
\def\eqsplit#1\ensplit{\begin{equation}\begin{split}#1\end{split}\end{equation}}
\def\alb#1\ale{\begin{align}#1\end{align}}
\def\nnb{\nonumber \\}            \def\nn{\nonumber}
\def\secl#1{\label{s:#1}}  \def\secr#1{Section~\ref{s:#1}}
\def\appl#1{\label{a:#1}}  \def\appr#1{Appendix~\ref{a:#1}}
\def\thml#1{\label{t:#1}}  \def\thmr#1{Theorem~\ref{t:#1}}
\def\prpl#1{\label{p:#1}}  \def\prpr#1{Proposition~\ref{p:#1}}
\def\leml#1{\label{l:#1}}  \def\lemr#1{Lemma~\ref{l:#1}}

\def\eql#1{\label{e:#1}}   \def\eqr#1{\eqref{e:#1}}
\newcounter{sequentialConst}
\def\Constn{\expandafter\ifx\csname Constcharacter\endcsname\relax%
\global\def\Constcharacter{C}%
\fi%
\stepcounter{sequentialConst}\Constcharacter_{\thesequentialConst}
}
\def\Constl#1{
\expandafter\xdef\csname Constno#1\endcsname{\thesequentialConst}
}
\def\Constr#1{\expandafter\ifx\csname Constno#1\endcsname\relax%
\typeout{Undefined ConstLABEL:  #1.}?\else%
\Constcharacter_{\csname Constno#1\endcsname}\fi}
\def\SPT#1{{\EuScript #1}}
\def\const{\mathchoice{\mbox{const.}}{\mbox{const.}}%
{\mbox{\scriptsize const.}}{\mbox{\tiny const.}}}
\newcommand{\reals}{{\mathbb R}}
\newcommand{\naturals}{{\mathbb N}}

\newcommand{\complexes}{{\mathbb C}}

\def\Tr{{\rm Tr\,}}
\def\supp{{\rm Supp\,}}

\def\Lim{\displaystyle\lim}
\def\Sum{\displaystyle\sum}

\def\blangle{\pmb{\langle}}
\def\brangle{\pmb{\rangle}}
\def\expt#1{\blangle #1 \brangle}
\def\lattice{\SPT{L}_\Lambda}
\def\vx{{\boldsymbol x}}
\def\vy{{\boldsymbol y}}
\def\vxi{{\boldsymbol\xi}}
\def\absx{|\vx|}
\def\trum#1#2{\mu_{#1,#2}^{(N)}}
\def\trun#1#2{\nu_{#1,#2}^{(N)}}
\def\trui#1#2{\nu_{#1,#2}^{(\infty)}}
\def\partition{{\cal P}}
\begin{document}
\title{
Triviality of hierarchical $O(N)$ spin model \\
in four dimensions with large $N$
}
\author{
Hiroshi Watanabe
     \\ {\small Department of Mathematics, Nippon Medical School,
   } \\ {\small 2--297--2, Kosugi, Nakahara, Kawasaki 211--0063, Japan
   } \\ {\small e-mail address: {\tt watmath@nms.ac.jp}
   }
}
\date{}
\maketitle
\begin{center}{\bf Abstract} \end{center}
{\small
The renormalization group transformation
for the hierarchical $O(N)$ spin model 
in four dimensions is studied
by means of characteristic functions of single-site measures,
and convergence of the critical trajectory to the Gaussian fixed point
is shown for a sufficiently large $N$.
In the strong coupling regime,
the trajectory is controlled by the help of the exactly solved $O(\infty)$ trajectory,
while, in the weak coupling regime,
convergence to the Gaussian fixed point is shown
by power decay of the effective coupling constant. 
}
\section{Introduction}\secl{intro}
In order to study a critical spin system 
with a large coupling constant,
it is necessary to control the renormalization group trajectory
in a strong coupling regime.
In the case of the hierarchical Ising model in four dimensions,
the method using characteristic functions of single--site measures was developed
and the critical trajectory was shown to converge to a Gaussian measure \cite{hhw}.
This means that the hierarchical Ising model in four dimensions is trivial,
namely, the continuum limit of the system is Gaussian.

In the present  paper,
we study the hierarchical $O(N)$ spin model in four dimensions 
and show the triviality of this model
for a sufficiently large $N$.


Let $N>1$ and $\Lambda>0$ be integers.
We consider Dyson's hierarchical spin model  \cite{dyson} 
with $O(N)$ symmetry
on the lattice $\lattice=\{0,1\}^\Lambda$:
\begin{align}\eql{phi}
  \phi_\theta&=\phi_{\theta_\Lambda,...,\theta_1}\in\reals^N\,, \quad
  \theta=(\theta_\Lambda,...,\theta_1)\in\lattice
\ ,\\
  H_\Lambda(\phi)&=-\dfrac12\sum_{n=1}^\Lambda\dfrac{1}{(2\omega)^n}
               \sum_{\theta_\Lambda,...,\theta_{n+1}=0,1}
               \left|\sum_{\theta_n,...,\theta_1=0,1}
               \phi_{\theta_\Lambda,...,\theta_1}\right|^2,
\\
  \langle F\rangle_{\Lambda,h_0^{(N)}} &=
 \dfrac1{Z_{\Lambda,h_0^{(N)}}}\int d\phi F(\phi)\exp(-\beta H_\Lambda(\phi))
                \prod_{\theta\in\lattice}h_0^{(N)}(\phi_\theta),
\\ \eql{Z}
  Z_{\Lambda,h_0^{(N)}} &= \int d\phi \exp(-\beta H_\Lambda(\phi))
            \prod_{\theta\in\lattice}h_0^{(N)}(\phi_\theta),
\end{align}
where $\beta>0$ and
\begin{align}
 \omega&=2^{2/d} \ ,\quad d>2\ .
\end{align}
For the normalized single site measure density $h_0^{(N)}$,
we choose
\begin{align}\eql{h0}
  h_0^{(N)}(\vx)=\const\delta(|\vx|-\sqrt{N}\alpha)
\ ,\quad \vx\in\reals^N\ ,
\end{align}
for $\alpha>0$.
This spin system is called {\it the $d$ dimensional hierarchical $O(N)$ spin model}
(slightly different from the version considered in \cite{gk2,gk4}). 
In what follows,
we shall fix the so far arbitrary normalization of the spin variables by 
\eqb  \eql{beta}
 \beta=\dfrac{\omega-1}{2}\ .
\eqe


Hierarchical models are so designed that
the block-spin renormalization group transformation ${\cal R}$ has a simple form.
Define {\it the block spins} $\phi'$ by
\[
  \phi'_\tau = 
  \dfrac{1}{\sqrt{2\omega}}\sum_{\theta_1=0,1}\phi_{\tau\theta_1}
\,,\  \tau=(\tau_{\Lambda-1},...,\tau_1). 
\]
If a function $F(\phi)$ depends on $\phi$ through $\phi'$ only,
namely, if there is a function $F'(\phi')$ of the block spins such that
 \[
  F(\phi)=F'(\phi'),
 \]
then it holds that
 \[
  \langle F\rangle_{\Lambda,h_0^{(N)}} 
  = \langle F'\rangle_{\Lambda-1,{\cal R}h_0^{(N)}}\,, 
 \]
where ${\cal R}$ is the mapping defined by
\begin{align}\eql{rg0}
  {\cal R}h(\vx)&=\const\exp(\dfrac\beta2 |\vx|^2)\int_{\reals^N}
 h(\sqrt{\dfrac{\omega}{2}}\vx+\vy)h(\sqrt{\dfrac{\omega}{2}}\vx-\vy)\,d\vy,\  
 \vx\in\reals^N.
\end{align}
Macroscopic properties for the spin system defined by \eqr{phi}--\eqr{Z}
are derived from the asymptotic behavior of the renormalization group trajectory
\begin{align}\eql{trajectory}
  h_n^{(N)}&={\cal R}^nh_0^{(N)}\ ,\quad n\ge0.
\end{align}


Note that
\begin{equation}
h_G(\vx) = \const \exp(-\dfrac{1}{4}|\vx|^2)
\end{equation}
is a fixed point of $\cal R$,
and the expectation $\langle\ \cdot\ \rangle_{\Lambda,h_G}$ defines a Gaussian measure.
We refer to $h_G$ as {\it the trivial fixed point} of $\cal R$.

In a weak coupling regime,
i.e. in a vicinity of the trivial fixed point $h_G$,
rigorous methods were developed to control 
the renormalization group trajectory for $\cal R$ \cite{gk2,gk4,sinai}.
However,
in order to show existence of the critical trajectory \eqr{trajectory},
we need 
to study the mapping $\cal R$
in the strong (as well as weak) coupling regime,
since the starting point \eqr{h0} is regarded as the strong coupling limit
of multi-component $\lambda\phi^4$ measure:
\[
 \delta(|\vx|-\gamma)
 =\lim_{\lambda\to\infty}\const\exp(-\lambda|\vx|^4+2\lambda\gamma^2|\vx|^2)
\ .
\]


In the present paper,
we study the trajectory \eqr{trajectory} 
and show that the critical trajectory converges to the trivial fixed point $h_G$.
To be precise:

\thmb\thml{main}
Let $d=4$.
For a sufficiently large $N$,
there exists a positive constant $\alpha_N$ such that
if $h_n^{(N)}, n\ge0,$ are defined by \eqr{h0} and \eqr{trajectory} with $\alpha=\alpha_N$,
then the sequence of measures $h_n^{(N)}(\vx)d\vx, n\ge0,$ 
weakly converges to the trivial fixed point measure  $h_G(\vx)d\vx$ as $n\to\infty$. 
\thme
 
As a result of the above theorem, we see that
`the continuum limit' constructed by using  the critical trajectory  is Gaussian.

Our proof of \thmr{main} is 
based on the method using characteristic functions of single--site measures 
developed in \cite{hhw}.
In the present paper, 
we analyze the $O(N)$ trajectories in the strong coupling regime
by explicitly solving $O(\infty)$ trajectories
and by estimating differences between $O(\infty)$ and $O(N)$ trajectories.
Thus, if $N$ is sufficiently large,
we can deal with the renormalization group transformation by hand
in contrast with the case $N=1$ solved in \cite{hhw},
in which the analysis in the strong coupling regime is partially computer aided.
On the other hand, 
our argument in the weak coupling regime is essentially the same as \cite{hhw}.

As is stated in \thmr{main},
we concentrate on the case $d=4$ and put
$\omega=\sqrt2$,
though a parallel argument is possible for $d>4$.
\section{Outline of the proof}
The proof of \thmr{main} is decomposed into three parts:
\begin{enumerate}
\item {\bf $O(N)$ trajectory in the weak coupling regime}
\par
We obtain a criterion
for the trajectory \eqr{trajectory} to converge to $h_G$
assuming that the trajectory has entered a vicinity of $h_G$
(\prpr{weak}).
Our criterion is stated in terms of characteristic functions
and differs from the one given in \cite{sinai},
which is so complicated that
it is not clear whether the trajectory starting at \eqr{h0} meets it.
\item {\bf $O(\infty)$ trajectory}
\par
We explicitly calculate the $O(\infty)$ trajectories,
i.e. the trajectories corresponding to  $N=\infty$,
and derive the asymptotic behavior of trajectories near the critical point (\prpr{Oinfty}).
\item {\bf From $O(N)$ trajectory to $O(\infty)$ trajectory}
\par
We show that an $O(N)$ trajectory converges to an $O(\infty)$ trajectory as $N\to\infty$
(\prpr{limit}).
Consequently, 
we can find the critical $O(N)$ trajectory
in the vicinity of the critical $O(\infty)$ trajectory
for a sufficiently large $N$.
\end{enumerate}

In this section,
we describe the outline of our argument
and prove \thmr{main}
assuming \prpr{weak}, \prpr{Oinfty} and \prpr{limit} stated below.
These propositions are proved in the subsequent sections.
\subsection{Characteristic functions}
We consider characteristic functions of effective measures
\[
 \hat h_n^{(N)}(\vxi)
={\cal F}h_n^{(N)}(\vxi)
=\int_{\reals^N} e^{\sqrt{-1}(\vxi,\vx)}h_n^{(N)}(\vx)d\vx
\ ,\quad n=0,1,2,\cdots,
\]
and write the renormalization group transformation for $\hat h_n^{(N)}$ as
\begin{equation}\eql{hierrecursion}
  \hat h_{n}^{(N)}={\cal F}{\cal R}{\cal F}^{-1}\hat h_{n-1}^{(N)}
   ={\cal T}{\cal S}\hat h_{n-1}^{(N)}  \,,
\end{equation}
where
\begin{align}
\eql{Strans} 
  {\cal S}g(\vxi)&=g(\dfrac{1}{\sqrt{2\omega}}\vxi)^2,
\\
\eql{Ttrans} 
  {\cal T}g(\vxi)&=\const \,\exp(-\dfrac{\beta}{2}\triangle)g(\vxi)\ .
\end{align}
In the above, $\triangle$ denotes the $N$ dimensional Laplacian
and the constant is chosen so that
\[
  {\cal T}g\,({\bf 0})=1
\]
holds.
Since $\hat h_n^{(N)}$ has spherical symmetry,
we shall often write
\[
 \hat h_n^{(N)}(\vxi)=\hat h_n^{(N)}(\xi),
\]
where $\xi=|\vxi|$.

The mapping $\cal T \cal S$ has the trivial fixed point $\hat h_G(\xi)=\exp(-\xi^2)$.

\subsection{The Lee--Yang property}
Let us introduce a `potential' $V_n^{(N)}(\xi)$
and its Taylor coefficients $\trum{k}{n}$ by
\begin{align}
\eql{dualpot}
\hat h_{n}^{(N)}(\xi) &=e^{-V_n^{(N)}(\xi)}\ ,
\\ 
\eql{trunptfcn}
V_n^{(N)}(\xi) &= \sum_{k=1}^\infty \trum{k}{n} \xi^{k}
\end{align}
for $n\ge0$.
(Note that $\hat h_{n}^{(N)}(0)=1$, i.e. $V_n^{(N)}(0)=0$.)
The coefficient $\trum{k}{n}$ is called {\it a truncated correlation}.
Since $\hat h_n^{(N)}(\xi)$ is even,
$\trum{k}{n}$ vanishes if $k$ is odd.

As is well-known,
the hierarchical model has {\it the Lee-Yang property} for any $N\ge1$:
$\hat h_n^{(N)}(\xi)$ has only  real zeros. (See e.g. \cite{K2}.)
As a result,
the truncated correlations have the bound
\cite{newman}:
\begin{equation}\eql{Taylorbound}
  0\le k \trum{2k}{n}\le (2\trum{4}{n})^{k/2},
  \ k\ge3, \ n\ge0.
\end{equation}
This implies the following:
\begin{enumerate}
\item
The Taylor expansion in the right hand side of \eqr{trunptfcn}
has a nonzero radius of convergence;
\item
It suffices to prove $\lim_{n\to\infty}\trum{4}{n}=0$ 
in order to ensure $\lim_{n\to\infty}\trum{2k}{n}=0$ for all $k\ge 2$,
which implies weak convergence of the trajectory to a Gaussian measure.
\end{enumerate}

Next we introduce {\it the scaled potential} $v_{n}^{(N)}(\eta)$
and its Taylor expansion by
\begin{align}\eql{scaledpotential}
 v_{n}^{(N)}(\eta)&=\dfrac{1}{N}V_{n}^{(N)}(\sqrt{N}\eta)
  = \sum_{k=1}^\infty \trun{k}{n} \eta^{k},\quad n\ge0.
\end{align}
In other words, we scale the truncated correlation  $\trum{k}{n}$ as
\[
 \trun{k}{n}=N^{k/2-1}\trum{k}{n},\quad k\ge1,\ n\ge0.
\]
Then, $\trun k n$ turns out to be ${\cal O}(1)$ with respect to $N$ (\lemr{conv}).
We refer to $\trun{k}{n}$ as {\it a scaled truncated correlation}.
In particular,
for the trivial fixed point measure $h_G(\vx)$,
the scaled potential is given by
\eqb\eql{trivialfp}
v_G(\eta)=\eta^2\ .
\eqe

\subsection{Differential equations for potentials}
In view of \eqr{Strans} and \eqr{Ttrans},
we consider the following equation:
\alb\eql{inverseheat}
 \dfrac{\partial}{\partial t}\hat h_{n}^{(N)}(t,\vxi)
 &=-\triangle\hat h_{n}^{(N)}(t,\vxi),
 \quad n\ge1,t\in[0,\beta/2]\ , 
\intertext{or, equivalently}
 \dfrac{\partial}{\partial t}\hat h_{n}^{(N)}(t,\xi)
 &=-\dfrac{\partial^2}{\partial\xi^2}\hat h_{n}^{(N)}(t,\xi)
   -\dfrac{N-1}{\xi}\dfrac{\partial}{\partial\xi}\hat h_{n}^{(N)}(t,\xi),
 \quad n\ge1,t\in[0,\beta/2]\ , \nn
\intertext{with the initial condition}
 \hat h_{n}^{(N)}(0,\xi)
 &=\hat h_{n-1}^{(N)}(\dfrac{1}{\sqrt{2\omega}}\xi)^2\ ,\quad n\ge1\ .\nn
\ale 
Then, we have
\[
 \hat h_{n}^{(N)}(\xi)
 =\dfrac{\hat h_{n}^{(N)}(\dfrac{\beta}{2},\xi)}{\hat h_{n}^{(N)}(\dfrac{\beta}{2},0)},
 \quad n\ge1\ .
\]
We also define the $t$-dependent scaled potential and its expansion by
\[
 v_n^{(N)}(t,\eta)=-\dfrac{1}{N}\log\hat h_n^{(N)}(t,\sqrt{N}\eta)
  = \sum_{k=1}^\infty \trun{k}{n}(t) \eta^{k}\ ,\quad n\ge1,t\in[0,\beta/2]\ .
\]
Then, the potentials $v_n^{(N)}(t,\eta), n\ge1,$ obey
\alb\eql{PDEuN}
 \dfrac{\partial}{\partial t}v_n^{(N)}(t,\eta) &= 
 \left({\dfrac{\partial}{\partial\eta}v_n^{(N)}(t,\eta)}\right)^2 
 -(1-\dfrac{1}{N})\dfrac{1}{\eta}\dfrac{\partial}{\partial\eta}v_n^{(N)}(t,\eta)
 - \dfrac{1}{N}\dfrac{\partial^2}{\partial\eta^2}v_n^{(N)}(t,\eta)
\ ,\\
 \eql{fromnu}
 v_n^{(N)}(0,\eta)&=2v_{n-1}^{(N)}(\dfrac{1}{\sqrt{2\omega}}\eta)
\ ,\\
 \eql{tonu+1}
 v_{n}^{(N)}(\eta)&=v_n^{(N)}(\dfrac{\beta}{2},\eta)-v_n^{(N)}(\dfrac{\beta}{2},0),
\intertext{and the Taylor coefficients $\nu_{2j,n}^{(N)}(t), j\ge1, n\ge1,$ obey}
\eql{ODEnu}
 \dfrac{d}{dt}\nu_{2j,n}^{(N)}(t)&=
 \mathop{\sum_{m+\ell=2j+2}}_{m,\ell\ge2}m\ell\nu_{m,n}^{(N)}(t)\nu_{\ell,n}^{(N)}(t)
  -(2j+2)(1+\dfrac{2j}{N})\nu_{2j+2,n}^{(N)}(t)
\ ,\\
\eql{initialnu}
 \nu_{2j,n}^{(N)}(0)&=\dfrac{2}{(2\omega)^{j}}\nu_{2j,n-1}^{(N)}
\ , \\
\eql{finalnu}
 \nu_{2j,n}^{(N)}&=\nu_{2j,n}^{(N)}(\dfrac{\beta}{2})\ .
\ale
The scaled potential \eqr{trivialfp} is a fixed point
of the above recursion relations.

Note that $\trun{2j}{n}(t)$ has the positivity due to the Lee--Yang property
\eqb\eql{LYt}
 \trun{2j}{n}(t)\ge 0\ ,\quad j\ge1, n\ge1\ ,
\eqe
since $\trun{2j}{n}(t)$ is regarded as a scaled truncated correlation
for a hierarchical model with $t$--dependence.
(See \eqr{phitheta}--\eqr{ZNLambda}.)

\subsection{Proof of \thmr{main}}\secl{proofmain}
In the weak coupling regime,
i.e. 
in the vicinity of the trivial fixed point \eqr{trivialfp},
we write $\trun{k}{n}, k=2,6,8,$
as follows:
\alb
 \eql{1zeta2}
 \nu_{2,n}^{(N)}&=
 1+\dfrac{1}{\sqrt2}(1+\frac2N)\nu_{4,n}^{(N)}+\zeta_{2,n}^{(N)}{\nu^{(N)}_{4,n}}^2
\ , \\
\eql{1zeta6}
 \nu_{6,n}^{(N)}&=4{\nu^{(N)}_{4,n}}^2+\zeta_{6,n}^{(N)}{\nu^{(N)}_{4,n}}^3
\ , \\
\eql{1zeta8}
 \nu_{8,n}^{(N)}&=\zeta_{8,n}^{(N)}{\nu^{(N)}_{4,n}}^3
\ ,
\ale
where $\trun{4}{n}$ is assumed to be small.
In fact, analyzing solutions to \eqr{ODEnu}--\eqr{finalnu},
we obtain the following proposition 
(proved at the end of \secr{weak}).
\prpb\prpl{weak}
Suppose that there exist 
a positive integer $n_1$ and positive constants $\alpha_\pm\ (\alpha_-<\alpha_+)$ such that
\begin{enumerate}
\item it holds that
\alb
 \zeta_{2,n_1}^{(N)}&=\zeta \ ,\quad \mbox{if\ } \alpha=\alpha_+\ ,\eql{zeta2+}\\
 \zeta_{2,n_1}^{(N)}&= -\zeta \ ,\quad \mbox{if\ } \alpha=\alpha_-\ ,\eql{zeta2-}
\ale
\item for $\alpha\in[\alpha_-,\alpha_+]$,
the following conditions are satisfied:
\alb
 |\zeta_{2,n_1}^{(N)}|&\le\zeta\ ,\nnb
 \nu_{4,n_1}^{(N)}&\le\epsilon\ ,\nnb
 |\zeta_{6,n_1}^{(N)}|\nu_{4,n_1}^{(N)}&\le\epsilon_0\ ,\nnb
 |\zeta_{8,n_1}^{(N)}|\nu_{4,n_1}^{(N)}&\le\epsilon_1\ ,\nn
\ale
\end{enumerate}
where $\zeta,\epsilon,\epsilon_0$ and $\epsilon_1$ are positive constants
determined in \secr{weakbounds}.
Then, there exists a value $\alpha_N\in[\alpha_-,\alpha_+]$
such that
\alb\eql{limit=1}
 \lim_{n\to\infty}\nu_{2,n}^{(N)}&=1
\  ,\\ \eql{limit=0}
 \lim_{n\to\infty}\nu_{4,n}^{(N)}&=0
\ale
hold at $\alpha=\alpha_N$.
\prpe
\bigskip\par
Next we formally put $N=\infty$ in \eqr{PDEuN}.
Namely, we consider the equation
\alb\eql{ODEnuinfty}
 \dfrac{\partial}{\partial t}v_n^{(\infty)}(t,\eta) &= 
 \left({\dfrac{\partial}{\partial\eta}v_n^{(\infty)}(t,\eta)}\right)^2 
 -\dfrac{1}{\eta}\dfrac{\partial}{\partial\eta}v_n^{(\infty)}(t,\eta)
\intertext{with}
 \eql{initialv}
 v_n^{(\infty)}(0,\eta)&=2v_{n-1}^{(\infty)}(\dfrac{1}{\sqrt{2\omega}}\eta)
\ ,\\  \eql{finalv}
 v_{n}^{(\infty)}(\eta)
 &=v_n^{(\infty)}(\dfrac{\beta}{2},\eta)-v_n^{(\infty)}(\dfrac{\beta}{2},0)\ ,
\ale
where the initial point is chosen as follows (see \lemr{inftyinitial}):
\eqb\eql{v0infty}
 v_{0}^{(\infty)}(\eta)
  =\int_0^\eta \dfrac{2\alpha^2\eta}{1+\sqrt{1-4\alpha^2\eta^2}}d\eta
  \left(=\lim_{N\to\infty} v_{0}^{(N)}(\eta)\right)\ .
\eqe
In \secr{Oinfty},
we solve \eqr{ODEnuinfty}--\eqr{v0infty}.
The solution is referred to as {\it the $O(\infty)$ trajectory}.
As is seen in \secr{Oinfty},
the critical value of $\alpha$ is $\sqrt{2+\sqrt2}$ 
and the critical trajectory tends to the trivial fixed point \eqr{trivialfp} 
as $n\to\infty$. 
(See \lemr{sp}.)

Now, consider the Taylor expansion
\eqb\eql{taylorinfty}
 v_n^{(\infty)}(\eta)=\sum_{j=1}^\infty \trui{2j}{n}\eta^{2j}
 \ ,\ n\ge0\ ,
\eqe
and write $\trui{k}{n},\ k=2,6,8,\ n\ge0$ as:
\alb
 \eql{0zeta2}
 \nu_{2,n}^{(\infty)}&=
 1+\dfrac{1}{\sqrt2}\nu_{4,n}^{(\infty)}+\zeta_{2,n}^{(\infty)}{\nu^{(\infty)}_{4,n}}^2
\ , \\
\eql{0zeta6}
 \nu_{6,n}^{(\infty)}&=4{\nu^{(\infty)}_{4,n}}^2+\zeta_{6,n}^{(\infty)}{\nu^{(\infty)}_{4,n}}^3
\ , \\
\eql{0zeta8}
 \nu_{8,n}^{(\infty)}&=\zeta_{8,n}^{(\infty)}{\nu^{(\infty)}_{4,n}}^3
\ .
\ale
Then we have the  following proposition (proved at the end of \secr{Oinfty}).
\prpb\prpl{Oinfty}
There exist a positive integer $n_1$ 
and positive constants $\alpha_{++},\alpha_{--}$ $(\alpha_{++}>\alpha_{--})$ 
such that
\begin{enumerate}
\item
it holds that
\alb\eql{Oinfty2+}
 \zeta_{2,n_1}^{(\infty)}&\ge2\zeta \ ,\quad\mbox{at } \alpha=\alpha_{++},\\
\eql{Oinfty2-}
 \zeta_{2,n_1}^{(\infty)}&\le-2\zeta \ ,\quad\mbox{at } \alpha=\alpha_{--},
\ale
\item
for $\alpha\in[\alpha_{--},\alpha_{++}]$,
the following conditions are satisfied:
\alb
\eql{Oinfty4}
 0<\nu_{4,n_1}^{(\infty)}&\le\dfrac12\epsilon\ ,\\
 |\zeta_{6,n_1}^{(\infty)}|\nu_{4,n_1}^{(\infty)}&\le\dfrac12\epsilon_0\ ,\\
\eql{Oinfty8}
 |\zeta_{8,n_1}^{(\infty)}|\nu_{4,n_1}^{(\infty)}&\le\dfrac12\epsilon_1\ .
\ale
\end{enumerate}
In the above,
$\zeta,\epsilon,\epsilon_0$ and $\epsilon_1$ are the same constants
as in \prpr{weak}.
\prpe
\bigskip\par
Finally we show that the $O(N)$ trajectory is approximated by the $O(\infty)$ trajectory
(proved at the end of \secr{fromN}).
\prpb\prpl{limit}
For each $j=1,2,\cdots,$ and for each $n=0,1,2,\cdots$,
it holds that
\eqb\eql{nunlimit}
 \lim_{N\to\infty}\trun{2j}{n}=\nu_{2j,n}^{(\infty)}\ .
\eqe 
The convergence is uniform in $\alpha$ on any compact subset of $(0,\infty)$.
\prpe

This fact is by no means  trivial,
because \eqr{PDEuN} is a singular perturbation of \eqr{ODEnuinfty},
to which the standard theory of differential equations does not apply:
note that \eqr{inverseheat} is a diffusion equation
in the inverse direction of time.
We show \prpr{limit} by means of $1/N$ expansion developed in \cite{K}.

\thmr{main} readily follows from
\prpr{weak},\prpr{Oinfty} and \prpr{limit}.

\prfofb{\thmr{main}}
We first use \prpr{Oinfty} and  fix the integer $n_1$.
Then using \prpr{limit} for $n=n_1$ and $j\le4$,
we see, for a sufficiently large $N$, that
\alb
 \zeta_{2,n_1}^{(N)}&\ge\zeta \ ,\quad\mbox{at } \alpha=\alpha_{++},\nnb
 \zeta_{2,n_1}^{(N)}&\le-\zeta \ ,\quad\mbox{at } \alpha=\alpha_{--},\nnb
\intertext{and that, for  $\alpha\in[\alpha_{--},\alpha_{++}]$,}
 0<\nu_{4,n_1}^{(N)}&\le\epsilon\ ,\nnb
 |\zeta_{6,n_1}^{(N)}|\nu_{4,n_1}^{(N)}&\le\epsilon_0\ ,\nnb
 |\zeta_{8,n_1}^{(N)}|\nu_{4,n_1}^{(N)}&\le\epsilon_1\ .\nn
\ale
Since $\zeta_{2,n_1}^{(N)}$ is continuous with respect to $\alpha\in[\alpha_{--},\alpha_{++}]$,
we can choose a subinterval $[\alpha_{-},\alpha_{+}]\subset[\alpha_{--},\alpha_{++}]$
so that the assumptions of \prpr{weak} are satisfied.
\thmr{main} follows from \eqr{limit=1} and  \eqr{limit=0} 
by virtue of \eqr{Taylorbound}.
\qed\prfofe

\section{$O(N)$ trajectory in weak coupling regime}\secl{weak}
In this section,
we analyze the solution to \eqr{ODEnu}--\eqr{finalnu}
in the weak coupling regime
and prove \prpr{weak}.
We shall abbreviate $\trun k n(t)$,  $\trun k n$ and $\zeta_{k,n}^{(N)}$
as $\nu_{k,n}(t)$,  $\nu_{k,n}$ and $\zeta_{k,n}$, respectively,
since we fix $N$ throughout this section.
All the bounds in this section are uniform in $N$.
\subsection{Recursion}
Let us consider \eqr{ODEnu}--\eqr{finalnu}.
We introduce functions $\lambda_{2j,n}(\tau), j\ge1, n\ge1,$ by
\[
 \nu_{2j,n}(t)=\sigma(t)\omega^{-1}\nu_{2,n-1}\delta_{j,1}
  +\sigma(t)^{2j}\lambda_{2j,n}(\sigma(t)t)
  \ ,\quad j\ge1, n\ge1,
\]
in order to separate the main contribution $\sigma(t)\omega^{-1}\nu_{2,n-1}$
to the `mass term' $\nu_{2,n}(t)$,
where
\[
\sigma(t)=\dfrac{1}{1-4\omega^{-1}\nu_{2,n-1}t}.
\]
In what follows,
we assume that
$\sigma(t)$ is defined for $t\in[0,\beta/2]=[0,(\omega-1)/4]$.
This is the case if $\nu_{2,n-1}$ is close to 1.

It is easily seen that
$\lambda_{2j,n}(\tau),\ j\ge1,$  satisfy the same equations 
as those for $\nu^{(N)}_{2j,n}(t)$:
\alb\eql{ODElambda}
 \dfrac{d}{d\tau}\lambda_{2j,n}(\tau)&=
 \mathop{\sum_{m+\ell=2j+2}}_{m,\ell\ge2}m\ell\lambda_{m,n}(\tau)\lambda_{\ell,n}(\tau)
  -(2j+2)(1+\dfrac{2j}{N})\lambda_{2j+2,n}(\tau)
\ ,\quad j\ge1\ ,\\
\intertext{with}
 \eql{initiallambda}
 \lambda_{2j,n}(0)&=\left\{
  \begin{array}{cl}
    0\ , & j=1\ , \\
    \dfrac{2}{(2\omega)^j}\nu_{2j,n-1}\ , & j\ge2\ ,
  \end{array} \right.\\
 \eql{finallambda}
 \nu_{2j,n}&=r\nu_{2,n-1}\delta_{j,1}+(r\omega)^{2j}\lambda_{2j,n}(T)
 \ ,\quad j\ge1\ ,
\intertext{where}
\eql{defr}
 r&=\dfrac{1}{1-(\omega-1)(\nu_{2,n-1}-1)}
\ ,\\
\eql{defT}
 T&=\dfrac14\omega(\omega-1)r \ .
\ale
Let us rewrite \eqr{ODElambda} and \eqr{initiallambda} for $j\le4$ as integral equations:
\alb\eql{intlam2}
 \lambda_{2,n}(\tau)&=\int_0^\tau(4\lambda_{2,n}(\tau)^2-4(1+\frac{2}{N})\lambda_{4,n}(\tau))d\tau
\ , \\
\eql{intlam4}
 \lambda_{4,n}(\tau)&=\dfrac{1}{2\omega^2}\nu_{4,n-1}
     +\int_0^\tau(16\lambda_{2,n}(\tau)\lambda_{4,n}(\tau)
       -6(1+\frac{4}{N})\lambda_{6,n}(\tau))d\tau
\ , \\
\eql{intlam6}
 \lambda_{6,n}(\tau)&=\dfrac{1}{4\omega^3}\nu_{6,n-1}
     +\int_0^\tau(24\lambda_{2,n}(\tau)\lambda_{6,n}(\tau)+16\lambda_{4,n}(\tau)^2
                  -8(1+\frac{6}{N})\lambda_{8,n}(\tau))d\tau
\ , \\
\eql{intlam8}
 \lambda_{8,n}(\tau)&=\dfrac{1}{8\omega^4}\nu_{8,n-1}
     +\int_0^\tau(32\lambda_{2,n}(\tau)\lambda_{8,n}(\tau)
            +48\lambda_{4,n}(\tau)\lambda_{6,n}(\tau)
                  -10(1+\frac{8}{N})\lambda_{10,n}(\tau))d\tau \ .
\ale

We now derive expressions for $\zeta_{k,n}, j=2,6,8,$ 
introduced in \eqr{1zeta2}--\eqr{1zeta8}
and confirm the `marginal behavior' of $\nu_{4,n}$
by using $\omega=\sqrt2$ ($d=4$).
Successive use of \eqr{intlam2}--\eqr{intlam8} yields
\alb
 \lambda_{2,n}(\tau)&=-4(1+\frac{2}{N})\dfrac14\nu_{4,n-1}\tau + X_2(\tau)
\ ,\nnb
 \lambda_{4,n}(\tau)&=\dfrac14\nu_{4,n-1}+X_4(\tau)
\ ,\nnb
 \lambda_{6,n}(\tau)&=\dfrac{1}{8\sqrt2}\nu_{6,n-1}+\nu_{4,n-1}^2\tau+X_6(\tau) 
\ ,\nnb
 \lambda_{8,n}(\tau)&=\dfrac{1}{32}\nu_{8,n-1}+X_8(\tau)\ ,\nn
\ale
where
\alb
 X_2(\tau)&=\int_0^\tau(4\lambda_{2,n}(\tau)^2-4(1+\frac{2}{N})X_4(\tau))d\tau\eql{defX2}\ ,\\
 X_4(\tau)&=\int_0^\tau(16\lambda_{2,n}(\tau)\lambda_{4,n}(\tau)
             -6(1+\frac{4}{N})\lambda_{6,n}(\tau))d\tau\eql{defX4}\ ,\\
 X_6(\tau)&=\int_0^\tau(24\lambda_{2,n}(\tau)\lambda_{6,n}(\tau)-8\nu_{4,n-1}X_4(\tau)
             +16X_4(\tau)^2
             -8(1+\frac{6}{N})\lambda_{8,n}(\tau))d\tau\eql{defX6}\ ,\\
 X_8(\tau)&=\int_0^\tau(32\lambda_{2,n}(\tau)\lambda_{8,n}(\tau)
    +48\lambda_{4,n}(\tau)\lambda_{6,n}(\tau)-10(1+\frac{8}{N})\lambda_{10,n}(\tau))d\tau
\ .\eql{defX8}
\ale
Using the above expressions, we obtain the following recursion relations:
\alb
 \zeta_{2,n}\nu_{4,n}^2&=\sqrt2r\zeta_{2,n-1}\nu_{4,n-1}^2+Y_2
   \eql{zeta2'}\ ,\\
 \zeta_{6,n}\nu_{4,n}^3&=\dfrac{1}{\sqrt2}r^6\zeta_{6,n-1}\nu_{4,n-1}^3+Y_6
   \eql{zeta6'}\ ,\\
 \zeta_{8,n}\nu_{4,n}^3&=\dfrac12r^8\zeta_{8,n-1}\nu_{4,n-1}^3+Y_8\ ,\\
 \nu_{4,n}&=\nu_{4,n-1}-(\dfrac12+\frac4N)\nu_{4,n-1}^2+Y_4 \ ,
   \eql{nu4'}
\ale
where
\alb
 Y_2&=\dfrac{1}{\sqrt2}(1+\frac2N)(\nu_{4,n-1}-\nu_{4,n})
      +\dfrac{r-1}{\sqrt2}(1+\frac2N)\nu_{4,n-1}
\nnb
&\mbox{\hspace{7mm}}
      +(1-\dfrac{1}{\sqrt2})(r-r^3)(1+\frac2N)\nu_{4,n-1}+2r^2X_2(T)\ , \\
 Y_4&=-(3-2\sqrt2)(1+\frac2N)(r^6-1)\nu_{4,n-1}^2
      -3(\sqrt2-1)(1+\frac{4}{N})(r^5-1)\nu_{4,n-1}^2
\nnb
&\mbox{\hspace{7mm}}
      -\dfrac{3(3-2\sqrt2)}{2}(1+\frac4N)(r^6-1)\nu_{4,n-1}^2
      +(4-2\sqrt2)(1+\frac2N)(r-1)\nu_{4,n-1}^2
\nnb
&\mbox{\hspace{14mm}}
      +4(\sqrt2-1)r\zeta_{2,n-1}\nu_{4,n-1}^3
      +(r-1)^2(r^2+2r+3)\nu_{4,n-1}+4r^4X_4(T)
\nnb
&\mbox{\hspace{21mm}}
      -64(1+\frac2N)r^4\nu_{4,n-1}\int_0^T\tau X_4(\tau)\ d\tau
      +64r^4\int_0^TX_2(\tau)\lambda_{4,n}(\tau)\ d\tau
\nnb
&\mbox{\hspace{28mm}}
      -\dfrac{3(\sqrt2-1)}{4}(1+\frac{4}{N})r^5\zeta_{6,n-1}\nu_{4,n-1}^3
      -24(1+\frac4N)r^4\int_0^TX_6(\tau)d\tau\ ,\\
 Y_6&=4(\nu_{4,n-1}^2-\nu_{4,n}^2)
      +2\sqrt2(r^6-1)\nu_{4,n-1}^2+(4-2\sqrt2)(r^7-1)\nu_{4,n-1}^2+8r^6X_6(T)\ ,\\
 Y_8&=16r^8X_8(T)\ .
\ale
\subsection{Bounds}\secl{weakbounds}
Let us derive bounds on $\zeta_{2j,n},\ j=1,3,4,$ and $\nu_{4,n}$
by means of \eqr{zeta2'}--\eqr{nu4'}.
Our starting point is:
\alb\eql{lam>0}
  \lambda_{k,n}(\tau)&\ge 0\ ,\quad k=4,6,8,10 \ ,\\
  \lambda_{2,n}(\tau)&\le 0\ . \eql{lam2<0}
\ale
The first inequality comes from \eqr{LYt}.
The second one is shown as follows.
{From} \eqr{ODEnu}, we see that $\nu_{2,n}(t)$ obeys
\[
 \dfrac{d}{dt}\nu_{2,n}(t)=4\nu_{2,n}(t)^2-4(1+\dfrac{2}{N})\nu_{4,n}(t)\ ,
\]
whereas the function $\tilde\nu_{2,n}(t)=\sigma(t)\omega^{-1}\nu_{2,n-1}$ satisfies
\[
 \dfrac{d}{dt}\tilde\nu_{2,n}(t)=4\tilde\nu_{2,n}(t)^2\ .
\]
Since $\nu_{2,n}(0)=\tilde\nu_{2,n}(0)$ and $\nu_{4,n}(t)\ge0$,
we have
\[
 \nu_{2,n}(t)\le \tilde\nu_{2,n}(t) \ ,
\]
and hence \eqr{lam2<0}.

Using \eqr{lam>0} and \eqr{lam2<0},
we obtain the following lemma.

\lemb\leml{zeta}
Suppose that  $\nu_{4,n-1}$ and $\zeta_{m,n-1}, m=2,6,8,$ satisfy
\alb
 \nu_{4,n-1}&\le\epsilon_0 , \eql{nu4<small}\\
 |\zeta_{m,n-1}|\nu_{4,n-1}&\le\epsilon_0\ ,\quad m=2,6,8. \eql{zetanu4<small}
\ale
Then $\nu_{4,n}$ and $\zeta_{m,n}, m=2,6,8,$ satisfy
\alb
 \nu_{4,n-1}-5\nu_{4,n-1}^2&<\nu_{4,n}<\nu_{4,n-1}-\dfrac14\nu_{4,n-1}^2,\eql{nu4bounds}\\
  |\zeta_{2,n}-\sqrt2\zeta_{2,n-1}|&<\Constn\Constl{zeta2N}
   \ , \eql{zeta2'lower}\\
 |\zeta_{6,n}|&<0.8 |\zeta_{6,n-1}|+\Constn(1+\zeta_{8,n-1}), \Constl{zetasix}
            \eql{zeta6'bound}\\
 |\zeta_{8,n}|&<0.6|\zeta_{8,n-1}|+\Constn\ , \Constl{zetaeight}
            \eql{zeta8'bound}
\ale 
where $\Constr{zeta2N}, \Constr{zetasix}, \Constr{zetaeight}$ and $\epsilon_0$
are positive constants independent of $N,n$ and $\alpha$.
\leme

\prfb
Suppose that $\nu_{k,n-1}, k=2,6,8,$ have the following forms
\alb
 \nu_{2,n-1}&=
 1+\dfrac{1}{\sqrt2}(1+\frac2N)\nu_{4,n-1}+\zeta_{2,n-1}\nu_{4,n-1}^2
\ , \nnb
 \nu_{6,n-1}&=4\nu_{4,n-1}^2+\zeta_{6,n-1}{\nu}_{4,n-1}^3
\ , \nnb
 \nu_{8,n-1}&=\zeta_{8,n-1}\nu_{4,n-1}^3
\nn
\ale
with the bounds \eqr{nu4<small} and \eqr{zetanu4<small}.
Then, if we choose $\epsilon_0$ sufficiently small,
equations \eqr{defr}--\eqr{intlam6} together with \eqr{lam>0} and \eqr{lam2<0} yield
\alb
& 1<r<1+5\nu_{4,n-1}
\ ,\nnb
& 0<T<\dfrac15
\ ,\nnb
& -\dfrac35\nu_{4,n-1}\le\lambda_{2,n}(\tau)\le0
\ ,\nnb
& 0\le\lambda_{4,n}(\tau)\le\dfrac14\nu_{4,n-1}
\ ,\nnb
& 0\le\lambda_{6,n}(\tau)\le\nu_{4,n-1}^2\ .\nn
\ale
Using the above bounds, we have for $\tau\in[0,T]$
\alb
 |X_2(\tau)|&\le\const\nu_{4,n-1}^2\ ,\nnb
 |X_4(\tau)|&\le\const\nu_{4,n-1}^2\ ,\nnb
 |X_6(\tau)|&\le\const(1+\zeta_{8,n-1})\nu_{4,n-1}^3\ ,\nnb
 X_8(\tau)&\le\const\nu_{4,n-1}^3\ ,\nn
\intertext{and hence}
 |Y_2|&\le\const\nu_{4,n-1}^2
\nnb
 |Y_4|&\le\const(1+|\zeta_{2,n-1}|+|\zeta_{6,n-1}|+|\zeta_{8,n-1}|)\nu_{4,n-1}^3
\nnb
 |Y_6|&\le\const(1+|\zeta_{8,n-1}|)\nu_{4,n-1}^3
\ ,\nnb
 Y_8&\le\const\nu_{4,n-1}^3\ .
\nn
\ale
The lemma follows from the above bounds and \eqr{zeta2'}--\eqr{nu4'}
combined with the assumptions \eqr{nu4<small} and \eqr{zetanu4<small},
if we choose appropriate constants $\Constr{zeta2N}, \Constr{zetasix}$ and $\Constr{zetaeight}$.
\qed\prfe

\remb
The function $\sigma(t)$ is well-defined on $[0,\beta/2]$
under \eqr{nu4<small} and \eqr{zetanu4<small},
if we choose $\epsilon_0$ sufficiently small.
\reme

Using the positive constants 
$\Constr{zeta2N}, \Constr{zetasix}, \Constr{zetaeight}$ and $\epsilon_0$
in \lemr{zeta},
we put 
\alb
 \zeta&=(\sqrt2+1)\Constr{zeta2N}
\ ,\\
 \epsilon_1&=\min(\epsilon_0,\dfrac{\epsilon_0}{10\Constr{zetasix}})
\ ,\\
 \epsilon&=
  \min(\epsilon_0 ,\dfrac{\epsilon_0 }{\zeta},\dfrac{\epsilon_0 }{10\Constr{zetasix}},
   \dfrac{2\epsilon_1}{5\Constr{zetaeight}})
\ .
\ale

\prfofb{\prpr{weak}}
Put $A=[\alpha_-,\alpha_+]$.
Under the assumption (2) of \prpr{weak},
we see that \eqr{nu4<small} and \eqr{zetanu4<small} are satisfied 
for $n=n_1+1$ and for any $\alpha \in A$.
{From} the assumption (1) and \eqr{zeta2'lower},
we have
\alb\eql{bs+}
 \zeta_{2,n_1+1}&>\zeta  \mbox{\hspace{27pt}at}\  \alpha=\alpha_+ \ , \\
\eql{bs-}
 \zeta_{2,n_1+1}&<-\zeta  \mbox{\hspace{20pt}at}\  \alpha=\alpha_-\ .
\ale
These bounds 
imply that
$\zeta_{2,n_1+1}$ runs through $[-\zeta,\zeta]$
when $\alpha$ scans $A$.
Then, we can find a subinterval $A'=[\alpha'_-,\alpha'_+]$ of $A$
such  that
\alb
 \zeta_{2,n_1+1}&=\pm\zeta  \mbox{\hspace{27pt}at}\  \alpha=\alpha'_\pm \ , \\
 |\zeta_{2,n_1+1}|&\le\zeta \mbox{\hspace{35pt}for}\  \alpha\in A' \ .
\ale 
Furthermore, from \eqr{nu4bounds},\eqr{zeta6'bound},\eqr{zeta8'bound} 
and the assumption (2) of \prpr{weak},
we have
\alb\eql{nu4to0}
 \nu_{4,n_1+1}&<\nu_{4,n_1}-\dfrac14\nu_{4,n_1}^2\ ,\\
 |\zeta_{6,n_1+1}|\nu_{4,n_1+1}&<\epsilon_0\ ,\\
 |\zeta_{8,n_1+1}|\nu_{4,n_1+1}&<\epsilon_1
\ale
for any $\alpha\in A$.
Thus, the assumptions of \prpr{weak} have been reproduced 
at the next stage
by replacing $\alpha_\pm$ by $\alpha'_\pm$.
Therefore, Bleher--Sinai argument \cite{BS1,BS2,sinai}\cite[P.25]{hhw} applies 
and,
\eqr{limit=1} and \eqr{limit=0} hold for some $\alpha\in A$
because of \eqr{nu4to0} and \eqr{1zeta2}.
\qed\prfofe
\section{$O(\infty)$ trajectory}\secl{Oinfty}
In this section, 
we solve the recursion relations \eqr{ODEnuinfty}--\eqr{finalv} with  \eqr{v0infty}.
Using the solution,
we derive bounds on the Taylor coefficients $\trui{2j}{n}, j=1,2,3,4,$
defined in \eqr{taylorinfty} and show  \prpr{Oinfty}.
\subsection{Recursions}
We firstly confirm the initial point \eqr{v0infty}.

\lemb\leml{inftyinitial}
Put
\[
 D_\theta=\{(\alpha,\eta)\in(0,\infty)\times\complexes\ |\ 2\alpha|\eta|<1
  \mbox{ or } |\arg\eta-\dfrac\pi2|<\theta
  \mbox{ or } |\arg\eta+\dfrac\pi2|<\theta\}
\]
for $\theta>0$.
Then, for a sufficiently small $\theta$, it holds that
\eqb\eql{inftyinitialpoint}
 \lim_{N\to\infty}\dfrac{d}{d\eta}v_{0}^{(N)}(\eta)
 =\dfrac{2\alpha^2\eta}{1+\sqrt{1-4\alpha^2\eta^2}}
 \ ,\quad (\alpha,\eta)\in D_\theta\ ,
\eqe
where the convergence is uniform 
in $(\alpha,\eta)$ on any compact subset of $D_\theta$.
Furthermore,
$\trun{2j}{0}, j\ge1,$ converges as $N\to\infty$.
\leme

\remb
In this section,
we use \eqr{inftyinitialpoint} only for $\eta$ with $2\alpha|\eta|<1$.
The result for $\eta$ near the imaginary axis will be used in \secr{2site}.
\reme

\prfb
The Fourier transform of \eqr{h0} is written as
\eqb\eql{Bessel}
 \hat h_0^{(N)}(\xi)
  =\Gamma(N/2)\dfrac{J_{N/2-1}(\sqrt{N}\alpha\xi)}{(\sqrt{N}\alpha\xi/2)^{N/2-1}}
\eqe
as a function of $\xi=|\vxi|$,
where $J_{\nu}$ denotes the Bessel function of order $\nu$:
\alb 
 J_{\nu}(z)
  &=\dfrac{\left(\dfrac{z}{2}\right)^{\nu}}{\sqrt{\pi}\Gamma(\nu+\frac12)}
   \int_0^\pi\cos(z\cos\theta)\sin^{2\nu}\theta d\theta\ .
\ale
The derivative of the scaled potential defined by \eqr{scaledpotential} is
therefore given by
\alb
 \dfrac{d}{d\eta}v_{0}^{(N)}(\eta)
  =-\dfrac{1}{N}
     \dfrac{\dfrac{d}{d\eta}\hat h_0^{(N)}(\sqrt{N}\eta)}{\hat h_0^{(N)}(\sqrt{N}\eta)}
  =\alpha\dfrac{J_{N/2}(N\alpha\eta)}{J_{N/2-1}(N\alpha\eta)}\ .
\ale  
Then, using the relation 
\[
 J_{\nu-1}(z) + J_{\nu+1}(z)=\dfrac{2\nu}{z}J_{\nu}(z)\ ,
\]
we have
\alb\eql{cfrac}
 \eta\dfrac{d}{d\eta}v_{0}^{(N)}(\eta)
&=\cfrac{c_0}{1-
 \cfrac{c_1}{1-
 \cfrac{c_2}{1-
 \cfrac{c_3}{1-
 \mbox{\raisebox{-10pt}{$\ddots$}}
 \mbox{\raisebox{-20pt}{$\dfrac{c_{\ell-1}}{1-w_\ell}$}} }}}}
 \ ,\quad \ell\ge1 \ ,
\ale
where
\alb\eql{c0}
 c_0&=\alpha^2\eta^2
\ ,\\
 \eql{cj}
 c_{j}&=\dfrac{\alpha^2\eta^2}{\left(1+\frac{2(j-1)}{N}\right)\left(1+\frac{2j}{N}\right)}
  \ ,\quad j\ge1
\ , \\
 \eql{w_ell}
 w_\ell&=\dfrac{\alpha\eta}{1+\frac{2(\ell-1)}{N}}
  \dfrac{J_{N/2+\ell}(\alpha N\eta)}{J_{N/2+\ell-1}(\alpha N\eta)}
 \ ,\quad \ell\ge1\ .
\ale

Now, 
let $\theta$ be a sufficiently small positive constant, 
and let $D'$ be any compact subset of $D_\theta$.
Then, 
by choosing a sufficiently large $\ell$ (depending on $D'$),
we see that
the right hand side of \eqr{cfrac} is well-defined for $(\alpha,\eta)\in D'$ 
and is holomorphic in $\eta$.
Furthermore, it holds that
\alb
 \lim_{N\to\infty}\eta\dfrac{d}{d\eta}v_{0}^{(N)}(\eta)
 =\cfrac{\alpha^2\eta^2}{1-
 \cfrac{{\alpha^2\eta^2}}{1-
 \cfrac{{\alpha^2\eta^2}}{1-
 \cfrac{{\alpha^2\eta^2}}{1-
 \mbox{\raisebox{-10pt}{$\ddots$}}}}}}
 =\dfrac{2\alpha^2\eta^2}{1+\sqrt{1-4\alpha^2\eta^2}}
\ale
for $(\alpha,\eta)\in D'$,
where the convergence is uniform on $D'$.
The last statement of the lemma is obvious.
\qed\prfe

Consider the recursion relations \eqr{ODEnuinfty}--\eqr{v0infty}.
Since $v_n^{(\infty)}(t,\eta)$  and $v_n^{(\infty)}(\eta)$ are even with respect to $\eta$,
we can define functions $u_n(t,x)$ and $u_n(x)$ by
\alb
 u_n(t,\eta^2)&=v_n^{(\infty)}(t,\eta)\ , \ n\ge1
\ ,\eql{defvninftyt}\\
 u_n(\eta^2)&=v_n^{(\infty)}(\eta)\ ,\ n\ge0
\ ,\eql{defvninfty}
\ale
respectively.
Then, \eqr{ODEnuinfty}--\eqr{v0infty} become
\alb\eql{eqforv}
\dfrac{\partial}{\partial t}u_n(t,x) 
&= 4x \left(\dfrac{\partial}{\partial x}u_n(t,x)\right)^2
-2\dfrac{\partial}{\partial x}u_n(t,x) 
\ ,\\
 u_n(0,x)&=2u_{n-1}(\dfrac{x}{2\omega})
\ , \eql{vinitial}\\
 u_{n}(x)&=u_n(\dfrac{\beta}{2},x)-u_n(\dfrac{\beta}{2},0)
\ , \eql{vfinal}\\
 u_0(x)&=\int_0^x\dfrac{\alpha^2}{1+\sqrt{1-4\alpha^2y}}dy
\ ,\eql{u'0}
\ale
where $n\ge1$.

Now, let us denote  the inverse of $p=u'_n(x)$  by $x=w_n(p)$
and the inverse of $p=u'_n(t,x)$  by $x=w_n(t,p)$ for each $t$.
Then, $w_n(p)$ and $w_n(t,p)$ obey the following recursion relations:
\alb
\eql{eqforw}
 \dfrac{\partial w_n}{\partial t}(t,p)
  &=-4p^2 \dfrac{\partial w_n}{\partial p}(t,p)-8pw_n(t,p)+2
\ ,\\
\eql{xpinitial2}
 w_n(0,p)&=2\omega w_{n-1}(\omega p) 
\ ,\\
 w_{n}(p)&=w_n(\dfrac{\beta}{2},p)
\ ,\eql{xpfinal}\\
 w_0(p)&=\dfrac{1}{2p}-\dfrac{\alpha^2}{4p^2}
\ .\eql{xp0}
\ale
\subsection{Solutions}
The equation \eqr{eqforw} with the initial condition \eqr{xpinitial2} 
can be explicitly solved.
In fact, as is checked by direct calculation,
the function
\alb\eql{2wp2att}
w_{n}(t,p)
 &=\dfrac{1}{2p^2}\left(p-\dfrac{1}{p^{-1}+4t}
 +\dfrac{4\omega}{(p^{-1}+4t)^2}w_{n-1}(\dfrac{\omega}{p^{-1}+4t})\right)
\ale
solves \eqr{eqforw} and \eqr{xpinitial2}.
This together with \eqr{xpfinal} and \eqr{beta} implies
\alb 
w_{n}(p)
 &=\dfrac{1}{2p^2}\left(p-\dfrac{1}{p^{-1}+\omega-1}
 +\dfrac{4\omega}{(p^{-1}+\omega-1)^2}w_{n-1}(\dfrac{\omega}{p^{-1}+\omega-1})\right)\ .
\ale
This recursion with \eqr{xp0} is then solved as follows:
\alb\eql{wnp}
 w_n(p)
& =\dfrac{1}{2p^2}
   \left(p+\dfrac12\sum_{j=1}^{n}\dfrac{2^{j}}{\omega^j-1+p^{-1}}
   -\dfrac12(\dfrac2\omega)^n\alpha^2\right)
 \ , \quad n\ge1, p>0\ .
\ale
Furthermore, this combined with \eqr{2wp2att} yields
\alb\eql{solwn}
 w_n(t,p)
 &=\dfrac{1}{2p^2}
   \left(p+\dfrac12\sum_{j=1}^{n}\dfrac{2^{j}}{\omega^j-\omega+4t+p^{-1}}
   -\dfrac12(\dfrac2\omega)^{n}\alpha^2\right)
 \ , \quad n\ge1, t\ge0, p>0\ .
\ale


Now, $u'_n(t,x)$ is obtained as the inverse of $x=w_n(t,p)$.
We here have to choose an appropriate branch of the multi-valued inverse.

\lemb\leml{x(s)}
\begin{enumerate}
\item
For $n\ge1$ and $t\ge0$, the equation
\eqb\eql{p0sol}
w_n(t,p)=0, p>0
\eqe
has a unique solution $p$,
which we hereafter denote by $\pi_{n}(t)$.
The function $\pi_n(t)$ is continuous with respect to $t\ge0$ 
and satisfies
\eqb\eql{der>0}
\dfrac{\partial w_n}{\partial p}(t,\pi_{n}(t))>0\ ,\quad t\ge0\ .
\eqe
\item
For $n\ge1$, 
there exists uniquely a continuous function $u'_n(t,x)$ defined on 
$\{(t,x)\in\reals\times\complexes\ |\ 
 t\in[0,\frac{\beta}{2}],\ |x|<\rho_n\}$ 
such that
\alb
 w_n(t,u'_n(t,x))&=x \ ,\quad  t\in[0,\frac{\beta}{2}],\ |x|<\rho_n
\ , \eql{wn=x}\\
 u'_n(t,0)&=\pi_{n}(t)\ ,\quad  t\in[0,\frac{\beta}{2}]
\ ,\eql{v'=p}
\ale
where $\rho_n>0$ is a suitably chosen constant independent of $t$.
Moreover, the functions $u'_n(t,x), n\ge1,$ are holomorphic with respect to $x$
and satisfy
\eqb
\eql{u'=omegau'}
\omega u'_{n+1}(0,2\omega x)=u'_{n}(\dfrac{\beta}{2},x)\ , \quad n\ge1\ ,
\  |x|<\min(\rho_n,\dfrac{\rho_{n+1}}{2\omega})\ ,
\eqe
\end{enumerate}
\leme

\prfb
\begin{enumerate}
\item
We see from \eqr{solwn} that 
$2p^2w_n(t,p)$ is increasing with respect to $p>0$ and 
\[
 2p^2w_n(t,p)\to \left\{
  \begin{array}{ll}
    -\dfrac12(\dfrac2\omega)^n\alpha^2\ , & p\to0\ ,\\
    +\infty\ , & p\to+\infty \ ,
  \end{array}\right.
\]
holds.
Then, \eqr{p0sol} has a unique solution $p$.
The remaining statements are obvious.
\item
Because of \eqr{der>0},
there exists uniquely a  continuous function $p=u'_n(t,x)$ defined on
$\{(t,x)\in\reals\times\complexes\ |\ 
 t\in[0,\frac{\beta}{2}],\ |x|<\rho_n\}$
satisfying \eqr{wn=x} and \eqr{v'=p} for some positive constant $\rho_n$
independent of $t$.
Since the function $w_n(t,p)$ is holomorphic on the complex half plane $\Re p>0$
for each  $t\in[0,\frac{\beta}{2}]$,
$u'_n(t,x)$ is holomorphic with respect to $x$.
Let us show \eqr{u'=omegau'}.
Since \eqr{wn=x},\eqr{xpinitial2} and \eqr{xpfinal} imply
\[
 x=w_n(0,u'_n(0,x))
   =2\omega w_{n-1}(\omega u'_n(0,x))
   =2\omega w_{n-1}(\dfrac{\beta}{2},\omega u'_n(0,x))\ ,
\]
we have
\eqb\eql{wn=wn-1}
  w_{n-1}(\dfrac{\beta}{2},\omega u'_n(0,2\omega x))=x\ .
\eqe
Furthermore, since \eqr{xpinitial2} and \eqr{xpfinal} imply
\[
 0=w_{n-1}(\dfrac{\beta}{2},\pi_{n-1}(\dfrac{\beta}{2}))
  =w_{n-1}(\pi_{n-1}(\dfrac{\beta}{2}))
  =\dfrac{1}{2\omega}w_{n}(0,\dfrac{1}{\omega}\pi_{n-1}(\dfrac{\beta}{2}))\ ,
\]
we have
\[
\dfrac{1}{\omega}\pi_{n-1}(\dfrac{\beta}{2})=\pi_{n}(0)\ ,
\]
that is,
\[
  \omega u'_n(0,0)=\omega \pi_{n}(0)=\pi_{n-1}(\dfrac{\beta}{2})\ .
\]
Therefore, we can conclude from \eqr{wn=wn-1} that 
\[
\omega u'_{n}(0,2\omega x)=u'_{n-1}(\dfrac{\beta}{2},x)
\]
holds
because of the uniqueness of $u'_{n-1}(\beta/2,x)$.
\end{enumerate}
\qed\prfe

\bigskip\par

Using the function $u'_n(t,x)$ in \lemr{x(s)},
we define $u_n(t,x)$ by
\eqb\eql{defun}
 u_n(t,x)
 =\int_0^{x}u'_n(t,y)dy-2\int_0^tu'_n(s,0)ds
\eqe
for $n\ge1, t\in[0,\frac{\beta}{2}]$ and for $x\in\complexes$ with $|x|<\rho_n$.
Obviously, it holds that
\eqb\eql{u'=grad}
 u'_n(t,x)=\dfrac{\partial}{\partial x}u_n(t,x)\ .
\eqe
We also define $u_n(x), n\ge0,$ so that \eqr{vinitial} holds.

\lemb\leml{inverse}
The functions $u_n(t,x), n\ge1,$ and $u_n(x), n\ge0,$ are 
holomorphic with respect to $x$ and satisfy \eqr{eqforv}--\eqr{u'0}.
Furthermore,
$p=u'_n(x)=\dfrac{d}{dx}u_n(x)$ is an inverse of $x=w_n(p)$.
\leme

\prfb
Equations \eqr{eqforw} and \eqr{wn=x} imply
\eqb\eql{eqforv'}
\dfrac{\partial}{\partial t}u_n'(t,x)
 =8xu'_n(t,x) \dfrac{\partial}{\partial x}u'_n(t,x)+4{u'_n(t,x)}^2
  -2\dfrac{\partial}{\partial x}u'_n(t,x)\ .
\eqe
Therefore, $u_n(t,x)$ satisfies \eqr{eqforv} because of \eqr{defun} and \eqr{u'=grad}.
Furthermore,  \eqr{vinitial}
and \eqr{u'=omegau'} imply
\[
\dfrac{\partial}{\partial x}u_n(x)=
\dfrac{\partial}{\partial x}u_n(\dfrac{\beta}{2},x),
\]
from which we obtain \eqr{vfinal}.

Next, we show \eqr{u'0}.
{From} \eqr{vinitial} with $n=1$, we see that
\[
 u'_0(x)=\omega u'_1(0,2\omega x).
\]
Then, \eqr{wn=x} and \eqr{xpinitial2} imply
\eqb\eql{x=w0u'0}
 x=\dfrac{1}{2\omega}w_1(0,u'_1(0,2\omega x))
  =w_0(\omega u'_1(0,2\omega x))
  =w_0(u'_0(x)).
\eqe
Because of \eqr{xp0}, this means
\eqb
 x=\dfrac{1}{2u'_0(x)}-\dfrac{\alpha^2}{4u'_0(x)^2}\ . \eql{xu'0}
\eqe
Since \eqr{xu'0} holds at $x=0$ as well,
we have \eqr{u'0}.

Finally, 
the function $p=u'_n(x)$ is an inverse of $x=w_n(p)$,
because $u'_n(x)=u'_n(\beta/2,x)$ and $w_n(p)=w_n(\beta/2,p)$.
\qed\prfe

Let us define $v_n^{(\infty)}(t,\eta)$ and $v_n^{(\infty)}(\eta)$ by
\eqr{defvninftyt} and \eqr{defvninfty}, respectively.
As a consequence of \lemr{inverse},
$v_n^{(\infty)}(\eta)$ and $v_n^{(\infty)}(t,\eta)$ 
can be Taylor-expanded as \eqr{taylorinfty} and
\eqb\eql{vntaylor}
 v_n^{(\infty)}(t,\eta)=\sum_{j=1}^\infty \trui{2j}{n}(t)\eta^{2j}\ ,
\eqe
respectively, in a neighborhood of $\eta=0$.
Thus, in view of \eqr{ODEnuinfty}--\eqr{finalv} and \eqr{v0infty},
we see that $\trui{2j}{n}(t)$ and $\trui{2j}{n}$ satisfy the following recursion relations:
\alb
\eql{ODEnui}
 \dfrac{d}{dt}\nu_{2j}^{(\infty)}(t)&=
 \mathop{\sum_{m+\ell=2j+2}}_{m,\ell\ge2}m\ell\nu_{m,n}^{(\infty)}(t)\nu_{\ell,n}^{(\infty)}(t)
  -(2j+2)\nu_{2j+2,n}^{(\infty)}(t)
\ , \\
\eql{initialnuinfty}
 \nu_{2j,n}^{(\infty)}(0)&=\dfrac{2}{(2\omega)^{j}}\nu_{2j,n-1}^{(\infty)}
\ , \\
\eql{finalnuinfty}
 \nu_{2j,n}^{(\infty)}&=\nu_{2j,n}^{(\infty)}(\dfrac{\beta}{2})
\ , \\
\eql{startnuinfty}
 \nu_{2j,0}^{(\infty)}&=\lim_{N\to\infty}\trun{2j}{0}\ .
\ale

\subsection{Asymptotics}
In what follows,
we derive the asymptotic forms \eqr{0zeta2}--\eqr{0zeta8}
with bounds on $\zeta_{k,n}^{(\infty)}, k=2,6,8,$
and show \prpr{Oinfty}.

As is shown in \lemr{inverse},
$p=u'_n(x)$ is an inverse of $x=w_n(p)$.
Then, the Taylor coefficient
\eqb\eql{nuinftyderp}
 \trui{2j}{n}=\dfrac{1}{j!}\dfrac{d^{j-1}p}{dx^{j-1}}(0)
\ ,\quad j\ge1\ ,\ n\ge1\ ,
\eqe
is calculated from \eqr{wnp}.
For convenience' sake,
we introduce the variable $s$ by
\eqb\eql{defofs}
 p=\dfrac{1}{1-s}\ 
\eqe
and regard $s$ as  a function of $x$.

\lemb\leml{sp}
The functions $s=s(x)$ and $p=p(x)$ satisfy
\alb\eql{equationfors}
 s&=\delta_n(2xp^2+\gamma_n-R_n(s))
\ale
where
\alb
 \delta_n&=\dfrac{1}{n/2+1}
\ ,\\
\eql{gamman}
 \gamma_n&=\dfrac{1}{\sqrt2}+(\alpha^2-2-\sqrt2)2^{n/2-1}
\ ,\\
 R_n(s)&=s^2(\dfrac{1}{1-s}+\dfrac{1}{2}\sum_{j=1}^{n}\dfrac{1}{2^{j/2}-s})\ .
\ale
\leme

\prfb
{From} \eqr{wnp}, we see that
\eqb\eql{wns}
 2xp^2=
   \dfrac{1}{1-s}+\dfrac12\sum_{j=1}^{n}\dfrac{2^{j}}{\omega^j-s}
   -\dfrac12(\dfrac2\omega)^n\alpha^2 \ .
\eqe
Substituting
\[
 \dfrac{2^j}{\omega^j-s}
 =\left(\dfrac{2}{\omega}\right)^j
  +\left(\dfrac{2}{\omega^2}\right)^js
  +\left(\dfrac{2}{\omega^3}\right)^j\dfrac{s^2}{1-\dfrac{s}{\omega^j}}
\ ,\quad j\ge0\ ,
\]
into \eqr{wns},
we obtain
\[
 2xp^2=-\gamma_n+\delta_n^{-1}s+R_n(s)\ ,
\]
where
\alb
 \gamma_n&=\dfrac{\omega-1}{2-\omega}
   +(\dfrac12\alpha^2-\dfrac{1}{2-\omega})\left(\dfrac{2}{\omega}\right)^n
\ ,\nnb
 \delta_n^{-1}&=1+\dfrac12\sum_{j=1}^n\left(\dfrac{2}{\omega^2}\right)^j
\ ,\nnb
 R_n(s)&=\dfrac{s^2}{1-s}
  +\dfrac12\sum_{j=1}^n\left(\dfrac{2}{\omega^3}\right)^j\dfrac{s^2}{1-\omega^{-j}s}
\ .\nn
\ale
Since $\omega=\sqrt2$, we have the lemma.
\qed\prfe

\remb
Eq.\eqr{gamman} implies that the critical value of $\alpha$ is $\sqrt{2+\sqrt2}$.
\reme

Consider the case $x=0$ in \eqr{equationfors}.
Then, $\sigma_n=s(0)$ satisfies
\alb\eql{equationfors0}
 \sigma_n&=\delta_n(\gamma_n-R_n(\sigma_n))\ ,
 \quad \sigma_n<1\ .
\ale

In \eqr{equationfors0},
if $\delta_n$ is small
and if $|\gamma_n|$ is not large,
then we see that $\sigma_n=\SPT{O}(\delta_n)$.
To be precise, 
using successive approximations,
we can show
\alb
 |\sigma_n|<\Constn\delta_n \Constl{sigman}
\ale
for $n$ and $\alpha$ satisfying e.g.
\alb
 \delta_n&<\dfrac{1}{10}, \\
 |\gamma_n|&<10\ ,   \eql{deltagamma2}
\ale
where $\Constr{sigman}$ is a positive constant.
\lemb\leml{zetainfty}
Suppose that $n$ and $\alpha$ satisfy \eqr{deltagamma2} and

\alb\eql{deltagamma1}
 n&>\Constn\Constl{n}\ .
\ale
Then, it holds that
\alb\eql{trui4bound}
\Constn\delta_n<\trui{4}{n}&<\Constn\delta_n
\Constl{nu4upper}
\ ,\\
\eql{trui4-delta}
|\trui{4}{n}-\delta_n|&<\Constn\delta_n^2
\ ,\\
\eql{zetabound}
|\zeta_{k,n}^{(\infty)}|&<\Constn  \Constl{zeta68}
\ ,\quad k=6,8.
\intertext{Furthermore, if we write}
\eql{defZn}
\zeta_{2,n}^{(\infty)}&=
 \dfrac{\delta_n}{{\trui{4}{n}}^2}(\gamma_n-\dfrac{1}{\sqrt2})+Z_n\ ,
\intertext{then $Z_n$ has the bound}
\eql{Zn}
|Z_n|&<\Constn\ .
\Constl{Zn}
\ale
In the above, $C_m$'s are positive constants independent of $n$.
\leme

\prfb
We choose the constant $\Constr{n}$ sufficiently large so that
\eqr{deltagamma2} and \eqr{deltagamma1} imply
\alb
 &|\sigma_n|<\Constr{sigman}\delta_n<\dfrac12 \ ,\nnb
 &\delta_n|R'_n(\sigma_n)|<\dfrac12\ . \nn  
\ale

Now, \eqr{defofs} and \eqr{equationfors} imply that
derivatives of $s=s(x)$ and of $p=p(x)$ are given by
\alb\eql{s'}
 s'(x)&=2\delta_np(x)^2D_n(x)
\ ,\\
\eql{p'}
 p'(x)&=p(x)^2s'(x)=2\delta_np(x)^4D_n(x)
\ ,
\ale
respectively, where
\eqb\eql{Dn}
D_n(x)=\dfrac{1}{1-4\delta_n xp(x)^3+\delta_n R'_n(s(x))}\ .
\eqe
Then, it holds that
\[
p'(0)
=\dfrac{2\delta_n}{(1-\sigma_n)^4(1+\delta_nR'_n(\sigma_n))}\ ,
\]
from which together with \eqr{nuinftyderp}
we obtain \eqr{trui4bound} and \eqr{trui4-delta}.
Furthermore, from \eqr{p'} we have
\alb\eql{p''}
 p''(x)&=8\delta_np(x)^3p'(x)D_n(x)+2\delta_np(x)^4D'_n(x)\ .
\ale
This implies
\[
 p''(0)-24\delta_n^2
   =8\delta_n(p(0)^3p'(0)D_n(0)-2\delta_n)
    +2\delta_n(p(0)^4D'_n(0)-4\delta_n)
\]
and hence
\[
 |p''(0)-24\delta_n^2|\le\const\delta_n^3 \ .
\]
Then, we have the bound \eqr{zetabound} on $\zeta_{6,n}^{(\infty)}$.
Similarly,
differentiating the both sides of \eqr{p''},
we obtain the bound on $\zeta_{8,n}^{(\infty)}$.

Let us show \eqr{Zn}.
Since \eqr{equationfors0} imply
\alb
 p(0)-1-\dfrac{1}{\sqrt2}\delta_n
  &=\sigma_n+\sigma_n^2p(0)-\dfrac{1}{\sqrt2}\delta_n
\nnb
  &=\delta_n(\gamma_n-\dfrac{1}{\sqrt2})-\delta_nR_n(\sigma_n)+\sigma_n^2p(0)
\ ,\nn
\intertext{we have}
\trui2n -1-\dfrac1{\sqrt2}\trui4n
  &=\delta_n(\gamma_n-\dfrac{1}{\sqrt2})
     -\delta_nR_n(\sigma_n)+\sigma_n^2p(0)
     +\dfrac1{\sqrt2}(\delta_n-\trui4n)
\ .\nn
\ale
This yields
\[
 Z_n=\dfrac{1}{{\trui4n}^2}(-\delta_nR_n(\sigma_n)+\sigma_n^2p(0)
     +\dfrac1{\sqrt2}(\delta_n-\trui4n))
\]
and hence the bound \eqr{Zn}.
\qed\prfe

\prfofb{\prpr{Oinfty}}
We choose $n_1$ sufficiently large so that \eqr{deltagamma1} and
\[
 \delta_{n_1}<\min\left( 
               \dfrac{1}{\Constr{nu4upper}^2(2\zeta+\Constr{Zn})},
               \dfrac{\epsilon}{2\Constr{nu4upper}},
               \dfrac{\epsilon_0}{2\Constr{nu4upper}\Constr{zeta68}},
               \dfrac{\epsilon_1}{2\Constr{nu4upper}\Constr{zeta68}},
               \right)
\]
hold.
By virtue of \eqr{gamman},
there exist $\alpha_{++}$ and $\alpha_{--}$ such that
\alb
 \gamma_{n_1}-\dfrac{1}{\sqrt2}&=\Constr{nu4upper}^2(2\zeta+\Constr{Zn})\delta_{n_1}
\ ,\quad \mbox{at } \alpha=\alpha_{++}
\ ,\nnb
 \gamma_{n_1}-\dfrac{1}{\sqrt2}&=-\Constr{nu4upper}^2(2\zeta+\Constr{Zn})\delta_{n_1}
\ ,\quad \mbox{at } \alpha=\alpha_{--}
\ .\nn
\ale
Since
\[
|\gamma_{n_1}-\dfrac{1}{\sqrt2}|\le1\ ,\quad \alpha\in[\alpha_{--},\alpha_{++}]\  ,
\]
\eqr{deltagamma2} holds for $n=n_1$ and $\alpha\in[\alpha_{--},\alpha_{++}]$.
Then, \eqr{trui4bound} and \eqr{zetabound} imply \eqr{Oinfty4}--\eqr{Oinfty8}.
Furthermore, \eqr{defZn} and \eqr{Zn} imply \eqr{Oinfty2+} and \eqr{Oinfty2-}.
\qed\prfofe
\section{From $\SPT{O}(N)$ trajectory to $\SPT{O}(\infty)$ trajectory}\secl{fromN}
In \secr{Oinfty}, we obtained the solution $\trui{k}{n}(t)$ 
to the system of ordinary differential equations 
\eqr{ODEnui}--\eqr{finalnuinfty} with \eqr{startnuinfty}.
This system is the formal limit of \eqr{ODEnu}--\eqr{finalnu} as $N\to\infty$.
In this section,
we show \prpr{limit},
namely, 
the fact that the solution $\trun{k}{n}(t)$ to \eqr{ODEnu}--\eqr{finalnu} is convergent
as $N\to\infty$
and the limit coincides with $\trui{k}{n}(t)$.

This fact is by no means  trivial,
because \eqr{PDEuN} is a singular perturbation of \eqr{ODEnuinfty}
and because the initial value problem for 
the infinite dimensional system of ordinary differential equations
lacks uniqueness of solution.
In order to show that $\trun{k}{n}(t)$ is convergent,
we use $1/N$ expansion developed by Kupiainen \cite{K}.
His method for spin systems on regular lattices
also applies to our hierarchical model.
\subsection{Boundedness}
%
%
We begin with the basic bound on the `mass term' $\trun2n(t)$ uniform in $N$.

\lemb
For $t\ge0, n\ge1, \alpha>0$ and $N\ge1$, it holds that
\alb\eql{nun2uniformbound}
 0\le\trun{2}{n}(t)\le\dfrac12\alpha^2\left(\dfrac2\omega\right)^n\ .
\ale
\leme

\prfb
We go back to the renormalization group transformation \eqr{rg0} in the configuration space:
\alb\eql{rechnN}
 h_n^{(N)}(t,\vx)&=\dfrac{1}{Z_n^{(N)}(t)}e^{t|\vx|^2}
 \int_{\reals^N} d\vy
 h_{n-1}^{(N)}(\sqrt{\dfrac{\omega}{2}}\vx+\vy)h_{n-1}^{(N)}(\sqrt{\dfrac{\omega}{2}}\vx-\vy)
\ ,
\intertext{where}
 Z_n^{(N)}(t)&=\int_{\reals^N}d\vx\ e^{t|\vx|^2}
 \int_{\reals^N} d\vy
 h_{n-1}^{(N)}(\sqrt{\dfrac{\omega}{2}}\vx+\vy)
 h_{n-1}^{(N)}(\sqrt{\dfrac{\omega}{2}}\vx-\vy)\ .
\ale
Note that $\trun{2}{n}(t)$ is related to the moment of the measure $h^{(N)}_n(t,\vx)d\vx$ as
\eqb\eql{trun2n}
 \trun{2}{n}(t)
      =\dfrac12\int_{\reals^N}x_1^2h_n^{(N)}(t,\vx)d\vx 
      =\dfrac{1}{2N}\int_{\reals^N}|\vx|^2h_n^{(N)}(t,\vx)d\vx\ ,
\eqe
where $\vx=(x_1,x_2,\cdots,x_N)$.

Let $r_n$ be the radius of $\supp h_n^{(N)}(\cdot)$ for $n\ge0$.
Since $\supp h_n^{(N)}(t,\cdot)=\supp h_n^{(N)}(\cdot)$ for $t\ge0$,
we see that $r_n$'s satisfy
\[
 r_n\le\sqrt{\dfrac{2}{\omega}}{r_{n-1}}\ ,\ n\ge1\ .
\]
Hence we have
\[
 r_n\le r_{0}\left(\dfrac2\omega\right)^{n/2}
        ={\sqrt{N}\alpha}\left(\dfrac2\omega\right)^{n/2}\ ,
\]
which together with \eqr{trun2n} implies \eqr{nun2uniformbound}.
\qed\prfe

Based on the above lemma,
we show the following bound independent of $N$.

\lemb\leml{nubound}
For $j\ge1, n\ge1$ and $\alpha_2>\alpha_1>0$,
there exists a positive constant
$\Constn\Constl{nujnNt}=\Constr{nujnNt}(2j,n,\alpha_1,\alpha_2)$
such that
\alb\eql{nun2juniformbound}
 0\le\trun{2j}{n}(t)\le\Constr{nujnNt}
\ale
holds for 
$\alpha\in[\alpha_1,\alpha_2],t\in[0,\beta/2]$ and $N\ge1$.
\leme

\prfb
For $n=0$, the lemma follows from \lemr{inftyinitial}.

Let $n\ge1$.
If $j=1$,
\eqr{nun2uniformbound} implies \eqr{nun2juniformbound}.
Let $j>1$.
We drop the last term in the right hand side of \eqr{ODEnu}
so that we obtain
\[
 \nu_{2j,n}^{(N)}(t)\le
 \dfrac{2}{(2\omega)^j}\trun{2j}{n-1}
 +\int_0^t\mathop{\sum_{m+\ell=2j+2}}_{m,\ell\ge2}
  m\ell\nu_{m,n}^{(N)}(\tau)\nu_{\ell,n}^{(N)}(\tau)d\tau\ .
\]
Then, we can show the lemma by induction on $n$ and $j$.
\qed\prfe
\subsection{Convergence}\secl{conv}
We now employ the $1/N$ expansion.
Let us consider the $\phi$ representation \eqr{phi}--\eqr{Z} in the following form:
\alb\eql{phitheta}
  \phi_\theta&=\phi_{\theta_\Lambda,...,\theta_1}\,,
   \ \  \theta=(\theta_\Lambda,...,\theta_1)\in\lattice
\ ,\\
  \eql{exp}
  \expt{F}_{\Lambda,t}&=
    \dfrac{1}{Z^{(N)}_\Lambda(t)}\int d\phi F(\phi)\exp(\frac12(\phi,J_\Lambda(t)\phi))
                \prod_{\theta\in\lattice}\delta(|\phi_\theta|-\sqrt{N}\alpha)
\ ,\\
 \eql{defJ}
  (\phi,J_\Lambda(t)\phi)&=
               \dfrac{2t}{(2\omega)^\Lambda}
               \left|\sum_{\theta_\Lambda,...,\theta_1=0,1}
               \phi_{\theta_\Lambda,...,\theta_1}\right|^2
               +
               \sum_{k=1}^{\Lambda-1}\dfrac{\beta}{(2\omega)^k}
               \sum_{\theta_\Lambda,...,\theta_{k+1}=0,1}
               \left|\sum_{\theta_k,...,\theta_1=0,1}
               \phi_{\theta_\Lambda,...,\theta_1}\right|^2
\ ,\\ \eql{ZNLambda}
  Z^{(N)}_\Lambda(t)&=\int d\phi \exp(\frac12(\phi,J_\Lambda(t)\phi))
            \prod_{\theta\in\lattice}\delta(|\phi_\theta|-\sqrt{N}\alpha)
\ .
\ale
Note that we have introduced $t$-dependence in the right hand side of \eqr{defJ}
in order to study the $t$-dependent correlation $\trun{k}{n}(t)$.

Let $\phi_\theta^{(i)}, i=1,2,\cdots,N,$ 
be the $i$-th component of $\phi_\theta\in\reals^N$.
For a set $A$ of lattice points in $\lattice$,
same point possibly occurring several times,
we write
\[
  \phi^{(i)}_{A}=\prod_{\theta\in A}\phi^{(i)}_\theta\ .
\]
Let
\eqb
 \expt{\prod_{i=1}^{N}\phi^{(i)}_{A_i}}_{\Lambda,t}
 =\sum_{m=0}^\infty s_m(\Lambda,t,\alpha,\{A_i\})N^{-m}
\eqe
be the formal $1/N$ expansion of 
the correlation $\expt{\prod_{i=1}^{N}\phi^{(i)}_{A_i}}_{\Lambda,t}$.
(See \cite[P.278]{K}.)

\prpb\prpl{K}{\rm\cite[Theorem 2]{K}}
For $r>0, \alpha_2>\alpha_1>0, \Lambda\ge2$,
there exists a positive constant
$\Constn\Constl{KN}=\Constr{KN}(r,\alpha_1,\alpha_2,\Lambda)$
such that, 
for $t\in[0,\beta/2], \alpha\in(\alpha_1,\alpha_2)$ and $N>\Constr{KN}$,
the remainder of the $1/N$ expansion up to $\SPT{O}(N^{-r+1})$ has the bound
\eqb
 \left|\expt{\prod_{i=1}^{N}\phi^{(i)}_{A_i}}_{\Lambda,t}
  -\sum_{m=0}^{r-1} s_m(\Lambda,t,\alpha,\{A_i\})N^{-m}\right|
 \le R(r,\alpha_1,\alpha_2,\Lambda,\{A_i\})N^{-r} \ ,
\eqe
where $R(r,\alpha_1,\alpha_2,\Lambda,\{A_i\})$ is a constant independent of $N$.
\prpe

\remb
The above proposition can be proved along the same line of argument as \cite{K}.
In \appr{hiergauss}, 
we check the properties to be assumed for applying the method of \cite{K} 
to our hierarchical system 
\eqr{phitheta}--\eqr{ZNLambda}.
Note also that $\Lambda\ge2$ should be assumed,
since we need bounds uniform in $t\in[0,\beta/2]$.
($J_1(t)$ vanishes as $t\to0$.)
\reme

\prpr{K} plays a key role in the proof of the lemma below (for $n\ge2$).
\lemb\leml{conv}
For $j\ge1$ and $n\ge1$,
the limit
\eqb
 \bar\nu_{2j,n}(t)=\lim_{N\to\infty}\trun{2j}{n}(t)
\eqe
exists,
where the convergence is uniform in $t\in[0,\beta/2]$
and in $\alpha$ on any compact subset of $(0,\infty)$.
\leme

\remb
In the proof below, 
we assume $n\ge2$ and apply \prpr{K}.
The case $n=1$ is dealt with in \secr{2site}.
\reme

\prfb
Let $n\ge2$.
Put, for $j\ge1$,
\alb\eql{xtoa}
 a_{2j,n}^{(N)}(t)&=\expt{x_1^{2j}}_{n,t}
\ ,\\
\eql{atonu}
a_{2j,n(c)}^{(N)}(t)&=(-1)^{j-1}\dfrac{(2j)!}{N^{j-1}} \nu^{(N)}_{2j,n}(t)
\ ,
\ale
where $x_1$ denotes the first component of the block spin
\eqb\eql{x1}
  \vx=\dfrac{1}{(2\omega)^{n/2}}
     \sum_{\theta_n,...,\theta_1}\phi_{\theta_n,...,\theta_1} \ .
\eqe
Note that $a_{2j,n(c)}^{(N)}(t)$ is the {\it connected part} of $a_{2j,n}^{(N)}(t)$,
namely,
\eqb\eql{xtoac}
 a_{2j,n(c)}^{(N)}(t)=\expt{x_1;x_1;\cdots;x_1}_{n,t}\ .
\eqe

Applying \prpr{K},
we expand 
correlations $a_{2j,n}^{(N)}(t)$ up to $\SPT{O}(N^{-j+1})$
with remainder estimates.
These expansions yield
an expression for the connected part $a_{2j,n(c)}^{(N)}(t)$,
in which all the terms up to $\SPT{O}(N^{-j+1})$ are explicitly written:
\eqb
 a_{2j,n(c)}^{(N)}(t)=\sum_{m=0}^{j-1}a_{2j,n,m(c)}(t)N^{-m}
     +a^{(N)\ge j}_{2j,n(c)}(t) \ ,
\eqe
where $a_{2j,n,m(c)}(t)N^{-m}$ stands for 
the sum of explicit terms of $\SPT{O}(N^{-m})$
and $a^{(N)\ge j}_{2j,n(c)}(t)$ denotes the remainder with the property:
\eqb
 \lim_{N\to\infty}N^{j-1}a^{(N)\ge j}_{2j,n(c)}(t)=0\ .
\eqe
Since \eqr{nun2juniformbound} implies that
the right hand side of \eqr{atonu} is uniformly bounded in $N$,
we see
\alb
 a_{2j,n,m(c)}(t)&=0\ ,\quad m=0,1,2,\cdots,j-2 \ ,
\intertext{so that we have}
 a^{(N)}_{2j,n(c)}(t)&=a_{2j,n,j-1(c)}(t)N^{-(j-1)}+a^{(N)\ge j}_{2j,n(c)}(t)\ .
\nn
\ale
This implies
\alb
 \trun{2j}{n}(t)=&\dfrac{(-1)^{j-1}}{(2j)!}a_{2j,n,j-1(c)}(t)
                +\dfrac{(-N)^{j-1}}{(2j)!}a^{(N)\ge j}_{2j,n(c)}(t) \nnb
              &\quad\to \dfrac{(-1)^{j-1}}{(2j)!}a_{2j,n,j-1(c)}(t) 
\ ,\quad N\to\infty \ .\nn
\ale
\qed\prfe
\subsection{Limit}
Thanks to \lemr{conv},
we can take limits in both sides of \eqr{ODEnu}--\eqr{finalnu} as  $N\to\infty$ :
\alb
\eql{ODEnuninftyA}
 \dfrac{d}{dt}{\bar\nu}_{2j,n}(t)&=
 \mathop{\sum_{m+\ell=2j}}_{m,\ell\ge2}m\ell\bar\nu_{m,n}(t)\bar\nu_{\ell,n}(t)
  -(2j+2)\bar\nu_{2j+2,n}(t)
\ ,\\
\eql{initialnuninftyA}
 \bar\nu_{2j,n}(0)&=\dfrac{2}{(2\omega)^{j}}{\bar\nu}_{2j,n-1}
\ ,\\
\eql{finalnuninftyA}
{\bar\nu}_{2j,n}&={\bar\nu}_{2j,n}(\dfrac{\beta}{2})\ ,
\intertext{and we can write \eqr{startnuinfty} as}
\eql{nubar=nuinfty}
{\bar\nu}_{2j,0}&=\trui{2j}{0}\ .
\ale

Although the system \eqr{ODEnuninftyA}--\eqr{finalnuninftyA} 
has the same form as \eqr{ODEnui}--\eqr{finalnuinfty},
we need an additional condition
in order to ensure the equality $\bar\nu_{2j,n}(t)=\trui{2j}{n}(t)$.

\lemb\leml{bar=infty}
In addition to  \eqr{ODEnui}--\eqr{finalnuinfty} and 
\eqr{ODEnuninftyA}--\eqr{nubar=nuinfty},
we assume
\eqb\eql{nu2=nu2}
 {\bar\nu}_{2,n}(t)={\nu}_{2,n}^{(\infty)}(t)
 \ ,\quad t\in[0,\dfrac{\beta}{2}], n\ge1.
\eqe
Then, it holds that
\eqb\eql{bar=infty}
{\bar\nu}_{2j,n}(t)={\nu}_{2j,n}^{(\infty)}(t)
 \ ,\quad t\in[0,\dfrac{\beta}{2}], n\ge1, j\ge1\ .
\eqe
\leme

\prfb
Solving \eqr{ODEnui} and \eqr{ODEnuninftyA} with respect to
$\nu_{2j+2,n}^{(\infty)}(t)$ and $\bar\nu_{2j+2,n}(t)$, respectively,
we see by induction that \eqr{bar=infty} holds.
\qed\prfe

\prfofb{\prpr{limit}}
Let us show \eqr{nu2=nu2}.
Since $p=u'_n(t,x)$ is an inverse of $x=w_n(t,x)$,
\eqr{defvninftyt} and \eqr{vntaylor} imply that
\eqb
\trui{2}{n}(t)=u'_n(t,0)=\pi_n(t)\ ,
\eqe
where $\pi_n(t)$ is the unique positive solution to $w_n(t,p)=0$ (see \lemr{x(s)}).
Then, in view of \eqr{solwn},
we see that $\trui{2}{n}(t)$ is the unique positive solution to
\eqb\eql{eqfornubar}
 \alpha^2=2(\dfrac{\omega}{2})^{n}\trui{2}{n}(t)+
   (\dfrac{\omega}{2})^{n}
   \sum_{j=1}^{n}\dfrac{2^{j}}{\omega^j-\omega+4t+\dfrac{1}{\trui{2}{n}(t)}}
\ .
\eqe

On the other hand,
using the variable $x_1$,
the first component of $\vx$  defined by \eqr{x1},
we have
\alb
{\bar\nu}_{2,n}(t)
&=\lim_{N\to\infty}\trun{2}{n}(t)
 =\lim_{N\to\infty}\dfrac12\expt{x_1^2}_{n,t}
\ ,\nnb
\eql{nuby2pf}
&=\lim_{N\to\infty}\dfrac{1}{2(2\omega)^n}
  \sum_{\theta,\theta'\in\SPT{L}_n}
    \expt{\phi^{(1)}_{\theta}\phi^{(1)}_{\theta'}}_{n,t}
\ .
\ale
In order to calculate
$\lim_{N\to\infty}\expt{\phi^{(1)}_{\theta}\phi^{(1)}_{\theta'}}_{n,t}$,
we use again the $1/N$ expansion.
(We can assume $n\ge2$ or $t>0$,
since, if $n=1$ and $t=0$, \eqr{nu2=nu2} has  been established in \eqr{nubar=nuinfty}.)
Decompose $J_n(t)$ defined by \eqr{defJ} with $\Lambda=n$ as follows:
\[
 J_n(t)=\triangle+\mu_0I \ .
\]
Here $I$ denotes the identity matrix and
the real constant $\mu_0$ is chosen so that 
\[
 \triangle{\bf 1}_n={\bf 0}
\]
holds, where we denoted ${\bf 1}_n={}^t(1,1,...,1)\in\reals^{\SPT{L}_n}$.

Now, put
\eqb
 C=(-\triangle+m^2I)^{-1}
\eqe
for $m^2>0$.

\lemb\leml{C}
It holds that
\alb\eql{csus}
\sum_{\theta,\theta'\in\SPT{L}_n}C_{\theta\theta'} &=\dfrac{2^n}{m^2}
\ ,\\ 
\eql{Cthetatheta}
C_{\theta\theta}&=
(\dfrac{\omega}{2})^n\sum_{j=1}^n
  \dfrac{2^j}{\omega^j-\omega+4t+2\omega^n m^2}+
  \dfrac{1}{2^n m^2}
\ ,\quad \theta\in\SPT{L}_n
\ .
\ale
\leme

This lemma will be shown in \appr{C}.

Now, choose $m^2$ so that 
$C_{\theta\theta}=\alpha^2$ holds, i.e.,
\eqb\eql{masschoice}
 (\dfrac{\omega}{2})^n\sum_{j=1}^n
  \dfrac{2^j}{\omega^j-\omega+4t+2\omega^n m^2}+
  \dfrac{1}{2^n m^2}=\alpha^2 \ .
\eqe
Then, we have \cite[P.284]{K}
\eqb\eql{lim=C}
\lim_{N\to\infty}\expt{\phi^{(1)}_{\theta}\phi^{(1)}_{\theta'}}_{n,t}
 =C_{\theta\theta'}\ .
\eqe

Now, \eqr{lim=C} together with \eqr{nuby2pf} and \eqr{csus} implies
\eqb\eql{Ctonuinfty}
{\bar\nu}_{2,n}(t)
=\dfrac12\dfrac{1}{(2\omega)^n}
  \sum_{\theta,\theta'\in\SPT{L}_n}C_{\theta\theta'}
  =\dfrac{1}{2\omega^n m^2} \ .
\eqe
Thus, 
we obtain \eqr{nu2=nu2}
from \eqr{masschoice}, \eqr{Ctonuinfty} and \eqr{eqfornubar}.

Since \eqr{nu2=nu2} holds,
\lemr{conv} and \lemr{bar=infty} yield \prpr{limit}.
\qed\prfofe

\section{The first effective theory}\secl{2site}
In this section, we prove \lemr{conv} for $n=1$.
For this purpose,
we study the first effective measure density:
\[
 h_1^{(N)}(t,\vx)=\dfrac{1}{Z_1^{(N)}(t)}\exp(t |\vx|^2)\int_{\reals^N}
   h_0^{(N)}(\sqrt{\dfrac{\omega}{2}}\vx+\vy)
   h_0^{(N)}(\sqrt{\dfrac{\omega}{2}}\vx-\vy)\,d\vy,\  \quad
 \vx\in\reals^N \ ,
\]
where
\eqb\eql{Z1}
 Z_1^{(N)}(t)= \int_{\reals^N}d\vx\exp(t\absx^2) \int_{\reals^N}d\vy
     h_0^{(N)}(\sqrt{\dfrac{\omega}{2}}\vx+\vy) 
     h_0^{(N)}(\sqrt{\dfrac{\omega}{2}}\vx-\vy) \ . 
\eqe

Let us denote integrations with respect to the measures 
$h_0^{(N)}(\vx)d\vx$ and $h_1^{(N)}(t,\vx)d\vx$ by
$\expt{\ \cdot\ }_0$ and $\expt{\ \cdot\ }_1$, respectively.

\lemb
Put
\eqb\eql{Am}
 A_m=\prod_{k=0}^{m-1}\left(1+\dfrac{2k}{N}\right)^{-1}
 \ ,\quad  m\ge1\ .
\eqe
Then, we have
\alb
\eql{h0byz^n}
 \hat h_0^{(N)}(\xi)&=
 1+\sum_{m=1}^{\infty}\dfrac{(-\alpha^2)^m}{2^m m!}A_m\xi^{2m}
\ ,\\
\eql{h1byz^n}
 \hat h_1^{(N)}(t,\xi)&=
 1+\sum_{m=1}^{\infty}\dfrac{(-1)^m}{(2N)^m m!}A_m\expt{\absx^{2m}}_1\xi^{2m}
\ .
\ale
\leme

\prfb
The equality \eqr{h0byz^n} follows from \eqr{Bessel}.
Let us show \eqr{h1byz^n}.

By virtue of spherical symmetry,
we have
\eqb\eql{h0h1}
 \hat h_1^{(N)}(t,\xi)
   =\expt{e^{\sqrt{-1}(\vxi,\vx)}}_1
   =\expt{e^{\sqrt{-1}\xi x_1}}_1
   =\sum_{m=0}^\infty\dfrac{(-\xi^2)^m}{(2m)!}\expt{x_1^{2m}}_1\ ,
\eqe
where $\xi=|\vxi|$ and $x_1$ is the first component of $\vx\in\reals^N$.
Similarly, we have 
\eqb\eql{h0expt0}
 \hat h_0^{(N)}(\xi)
   =\sum_{m=0}^\infty\dfrac{(-\xi^2)^m}{(2m)!}\expt{x_1^{2m}}_0\ .
\eqe
Comparing \eqr{h0expt0} with \eqr{h0byz^n}, 
we see that
\eqb\eql{h0h0}
 \expt{x_1^{2m}}_0=\dfrac{\alpha^{2m}(2m)!}{2^m m!}A_m
   =\dfrac{(2m)!}{(2N)^m m!}A_m\expt{\absx^{2m}}_0 \ .
\eqe
Now, note that the expectation $\expt{\ \cdot\ }_0$
gives average over an $N$ dimensional sphere.
Then,
integrating with respect to the radial distribution of $\hat h_1^{(N)}(t,\vxi)$,
we obtain from \eqr{h0h0}
\[
 \expt{x_1^{2m}}_1
   =\dfrac{(2m)!}{(2N)^m m!}A_m\expt{\absx^{2m}}_1 \ .
\]
This together with \eqr{h0h1} gives \eqr{h1byz^n}.
\qed\prfe

Now let us introduce {\it the connected part}
\eqb\eql{x2k1cc}
\expt{|\vx|^{2k}}_{1,cc}=\langle\absx^2;\absx^2;\cdots;\absx^2\rangle_1
\eqe
of  $\expt{|\vx|^{2k}}_1$ 
with $|\vx|^{2}$ as units, namely,
we define $\expt{|\vx|^{2k}}_{1,cc}, k=1,2,\cdots,$ by
\eqb
 \sum_{m=0}^\infty \dfrac{\expt{|\vx|^{2m}}_1}{m!}z^m
 =\exp\left(\sum_{k=1}^\infty \dfrac{\expt{|\vx|^{2k}}_{1,cc}}{k!}z^k\right)\ ,
\eqe
or, equivalently
\alb\eql{1to1cc}
\expt{\absx^{2m}}_1
&=\sum_{n\ge1}\sum_{\sum_{j=1}^{n} k_j=m}
\dfrac{1}{n!}\dfrac{m!}{\prod_{j=1}^n k_j!}
\ \prod_{j=1}^n\expt{\absx^{2k_j}}_{1,cc}\ ,\quad m\ge1\ ,
\ale
where the summations are taken over the sets of all 
$(k_1,k_1,\cdots,k_n)\in\naturals^n$ 
with $\Sum_{j=1}^n k_j=m$
and over $n\in\naturals$.
Note that we can get information on 
the connected part $\expt{\absx^{2k}}_{1,cc}$
from $Z_1^{(N)}(t)$, since
\eqb\eql{betaderivative}
 \expt{\absx^{2k}}_{1,cc}=\left(\dfrac{d}{dt}\right)^k\log Z_1^{(N)}(t)\ ,
 \quad k\ge1\ .
\eqe

\lemb\leml{Zlimit}
For $k=1,2,\cdots$,
the limit $\Lim_{N\to\infty}\dfrac{1}{N}\expt{\absx^{2k}}_{1,cc}$ exists,
where the convergence is uniform in $t\in[0,\beta/2]$ 
and in $\alpha$ on any compact subset of $(0,\infty)$.
\leme

\prfb
We carry out the integrations in the right hand side of \eqr{Z1}:
\alb
 Z_1^{(N)}(t)
     &=\const\int_{\reals^N}d\vx\int_{\reals^N}d\vy
       \exp(\frac{t}{2\omega}|\vx+\vy|^2)
       h_0^{(N)}(\vx)h_0^{(N)}(\vy) \nnb
     &=\const \exp(\dfrac{t}{\omega}N\alpha^2)\int_{\reals^N}d\vx\int_{\reals^N}d\vy
       \exp(\frac{t}{\omega}(\vx,\vy))
       \delta(|\vx|-\sqrt{N}\alpha)
       \delta(|\vy|-\sqrt{N}\alpha)\nnb
     &=\const \exp(\dfrac{t}{\omega}N\alpha^2)\int_{\reals^N}d\vx
       \delta(|\vx|-\sqrt{N}\alpha)
       \hat h_0^{(N)}(\frac{t}{\sqrt{-1}\omega}|\vx|)\nnb
     &=\const \exp(\dfrac{t}{\omega}N\alpha^2)
       \hat h_0^{(N)}(\frac{t}{\sqrt{-1}\omega}\sqrt{N}\alpha)\ . \nn
\ale
Therefore, we have
\alb
\eql{logZ}
 \dfrac{1}{N}\log Z_1^{(N)}(t)
     &=\const+\dfrac{t}{\omega}\alpha^2
        -v_0^{(N)}(\frac{t\alpha}{\sqrt{-1}\omega}) \ .
\ale
\lemr{inftyinitial} implies that
the right hand side of \eqr{logZ} converges as $N\to\infty$,
where the convergence is uniform in $t$ 
on any compact subset of a certain neighborhood of the real axis
and in $\alpha$ on any compact subset of $(0,\infty)$.
Then, the limit
\[
 \lim_{N\to\infty}\dfrac{1}{N}\expt{\absx^{2k}}_{1,cc}
= \lim_{N\to\infty}\dfrac{1}{N}
  \left(\dfrac{d}{dt}\right)^k\log Z_1^{(N)}(t)
\]
exists for $k\ge1$, $t\in[0,\beta/2]$ and for $\alpha>0$.
The convergence is uniform in $t\in[0,\beta/2]$ and in $\alpha$ on any compact subset
of $(0,\infty)$.
\qed\prfe

Next, we define the connected part $A_{k,c}$ of $A_k$ by
\eqb\eql{1Akc}
 1+\sum_{m=1}^\infty \dfrac{A_m}{m!}z^m
 =\exp\left(\sum_{k=1}^\infty \dfrac{A_{k,c}}{k!}z^k\right)\ ,
\eqe
or equivalently
\alb\eql{Akc}
A_{m}
&=\sum_{n\ge1}\sum_{\sum_{j=1}^{n} k_j=m}
\dfrac{1}{n!}\dfrac{m!}{\prod_{j=1}^n k_j!}
\ \prod_{j=1}^n A_{k_j,c}
\ ,\quad m\ge1\ .
\ale

\lemb\leml{NAlimit}
For $k=1,2,\cdots$,
the limit $\Lim_{N\to\infty}N^{k-1}A_{k,c}$ exists.
\leme

\prfb
{From} \eqr{h0byz^n} and \eqr{1Akc}, we have
\alb
 \hat h_0^{(N)}(\xi)
 &=\exp\left(\sum_{k=1}^\infty \dfrac{A_{k,c}}{k!}(-\dfrac12\alpha^2\xi^2)^{k}\right)
\nn
\ale
and hence 
\alb
 v_0^{(N)}(\eta)
  &=-\dfrac{1}{N}\sum_{k=1}^\infty \dfrac{A_{k,c}}{k!}(-\dfrac12\alpha^2 N\eta^2)^{k} \ .
\nn
\ale
Then, it holds that
\[
 N^{k-1}A_{k,c}=(-1)^{k-1}\dfrac{k!2^k}{\alpha^{2k}}\trun{2k}{0} \ .
\]
Since \lemr{inftyinitial} implies that
$\trun{2k}{0}$ converges as $N\to\infty$,
we obtain the lemma.
\qed\prfe

Put, for $m\ge1$,
\alb
 B_m&=\expt{\absx^{2m}}_1
\ ,\nnb
 B_{m,c}&=\expt{\absx^{2m}}_{1,cc}
\ ,\nn
\ale
and  write \eqr{h1byz^n} as
\eqb
\hat h_1^{(N)}(\xi)
=\sum_{m=1}^{\infty}\dfrac{A_mB_m}{m!}(-\dfrac{\xi^2}{2N})^{m} \ .
\eql{h1AB}
\eqe

We now introduce connected parts of the product ``$AB$''.
Let ${\cal P}_m$ be the set of all partitions of $\{1,2,\cdots,m\}$.
Then, \eqr{Akc} and \eqr{1to1cc} can be written as
\alb
 A_m&=\sum_{P\in\partition_m}\prod_{I\in P}A_{|I|,c}\ ,\nnb
 B_m&=\sum_{P\in\partition_m}\prod_{I\in P}B_{|I|,c}\ ,\nn
\ale
respectively, 
where $|I|$ denotes the number of elements of $I$.
Here, we think each $I\in\partition_m$ to connect elements contained in $I$ 
into a single component
by $|I|-1$ links (bonds).
The connected part of ``$AB$'' is by definition
\eqb\eql{ABc}
 (AB)_{k,c}=
 \sum_{P,Q\in\partition_k}\chi_k(P,Q)\prod_{I\in P}A_{|I|,c}\prod_{J\in Q}B_{|J|,c}
\ ,
\eqe
where
\eqb\eql{chiPQ}
 \chi_k(P,Q)=\left\{\begin{array}{ll}
   1, & \{1,2,\cdots,k\} \mbox{ is connected by } P\cup Q\ ,\\
   0, & \mbox{otherwise}\ .
 \end{array}\right.
\eqe
In this notation, we can write \eqr{h1AB} as
\[
 \hat h_1^{(N)}(\xi)
 =\exp\left(\sum_{k=1}^\infty \dfrac{(AB)_{k,c}}{k!}\left(-\dfrac{\xi^2}{2N}\right)^k\right)\ .
\]
Thus, we obtain the formula:
\eqb\eql{nu=AB}
 \trun{2k}{1}(t)=-\dfrac{1}{N}(-\dfrac{1}{2})^k\dfrac{(AB)_{k,c}}{k!} \ .
\eqe

\lemb\leml{limAB}
For $k=1,2,3,\cdots$,
the limit $\Lim_{N\to\infty}\dfrac1N(AB)_{k,c}$ exists,
where the convergence is uniform in $t\in[0,\beta/2]$ 
and in $\alpha$ on any compact subset of $(0,\infty)$.
\leme

\prfb
Consider the term
\[
 \psi(P,Q)=\prod_{I\in P}A_{|I|,c}\prod_{J\in Q}B_{|J|,c}
\]
in the right hand side of \eqr{ABc}
corresponding to a pair of partitions $P,Q\in\partition_k$ with $\chi_k(P,Q)=1$.
Suppose that the partition $P$ consists of $n$ elements,
that is, the set $\{1,2,\cdots,k\}$ is decomposed into $n$ disjoint subsets
each of which is connected by some element of $P$.
Then,  \lemr{NAlimit} implies that the quantity
\[
N^{k-n}\prod_{I\in P}A_{|I|,c}= \prod_{I\in P}N^{|I|-1}A_{|I|,c}
\]
is convergent as $N\to\infty$.
On the other hand, $Q$ should have at least $n-1$ links
because otherwise the set $\{1,2,\cdots,k\}$ could not be connected by $P\cup Q$.
Let us denote the number of elements of $Q$ by $n'$.
Then, we have $n'\le k-n+1$ and \lemr{Zlimit} implies that
\[
 N^{-k+n-1}\prod_{J\in Q}B_{|J|,c}
= N^{n'-(k-n+1)}\prod_{J\in Q}N^{-1}B_{|J|,c}
\]
is convergent.
Thus, we see that
\[
 \dfrac{1}{N}\psi(P,Q)
 =N^{k-n}\prod_{I\in P}A_{|I|,c} \cdot N^{-k+n-1}\prod_{J\in Q}B_{|J|,c}
\]
is convergent and
we obtain the lemma.
\qed\prfe

\lemr{conv} for $n=1$ directly follows from \lemr{limAB} and \eqr{nu=AB}.

\bigskip\bigskip\par\noindent
{\it Acknowledgments.\ }
I would like to thank T.Hattori and T.Hara
for valuable discussions and comments.
\bigskip\bigskip\par

\appendix
\section{Hierarchical Gaussian}\appl{hiergauss}
In this appendix,
we show the properties to be assumed 
when we  apply the $1/N$ expansion \cite{K} to the proof of \prpr{K}:
we prove the reflection positivity (\lemr{RP}) and 
certain bounds on the hierarchical Laplacian (\lemr{choiceofm2}).
In addition we prove \lemr{C}.

Let us write \eqr{phitheta}--\eqr{ZNLambda} as follows:
\alb
  \phi_\theta&=\phi_{\theta_\Lambda,...,\theta_1}\,,
   \ \  \theta=(\theta_\Lambda,...,\theta_1)\in\lattice
\ ,\\
  \expt{F}_{J}&=
    \dfrac{1}{Z_J}\int d\phi F(\phi)\exp(\frac12(\phi,J\phi))
                \prod_{\theta\in\lattice}\delta(|\phi_\theta|^2-N\alpha^2)
\ ,\\
  (\phi,J\phi)&=
               \sum_{k=1}^{\Lambda}
               \dfrac{\beta_k}{(2\omega)^k}
               \sum_{\theta_\Lambda,...,\theta_{k+1}=0,1}
               \left|\sum_{\theta_k,...,\theta_1=0,1}
               \phi_{\theta_\Lambda,...,\theta_1}\right|^2
\ ,\\
  Z_J(t)&=\int d\phi \exp(\frac12(\phi,J\phi))
            \prod_{\theta\in\lattice}\delta(|\phi_\theta|^2-N\alpha^2)
\ ,
\ale
where 
\eqb\eql{betak}
 \beta_k=\left\{
  \begin{array}{ll}
    \beta\ ,& k=1,2,\cdots,\Lambda-1 \ , \\ 2t \ ,& k=\Lambda \ .
  \end{array}\right.
\eqe
We define a matrix $B$ by
\eqb\eql{BST}
 (B\phi)_\tau=\dfrac{1}{\sqrt2}\sum_{\theta_1=0,1}\phi_{\tau\theta_1}
\eqe
and write
\eqb\eql{JbyBB}
 J=\sum_{k=1}^{\Lambda}\dfrac{\beta_k}{\omega^k}{B^*}^{k}B^k\ .
\eqe
Precisely saying,
$B^k$ should be written as a product of distinct matrices $B_1,B_2,\cdots,B_k$.
We however suppressed the subscripts of $B$ for simplicity.
\subsection{Reflection Positivity}
For $l=1,2,\cdots,\Lambda$,
we define the reflection $\rho_l$ on the lattice $\lattice$ by
\[
 (\rho_l\theta)_k=\left\{\begin{array}{ll}
      \theta_k, & k\neq l \ ,\\
      1-\theta_k, & k=l \ , \\
      \end{array}\right.
\quad \theta\in\lattice\ .
\]

\lemb\leml{RP}
The measure $\expt{\cdot}_{J}$ has reflection positivity 
with respect to $\rho_l, l=1,2,\cdots,\Lambda$.
\leme

\remb
Since the reflection planes for $\rho_l, l=1,2,\cdots,\Lambda,$
separate the $2^\Lambda$ points in $\lattice$ from each other,
we have ``the chessboard bound'' \cite{FILS} used in \cite{K} from this lemma.
\reme

\prfb
We fix $l$.
Let us define the upper half space with respect to $\rho_l$ by
\[
 \lattice^+=\{\theta\ |\ \theta_l=1\}
\]
and denote the set of all polynomials in
$\phi_\theta, \theta\in\lattice^+,$
by  ${\bf P^+}$.
The lower half space $\lattice^{-}$ and 
the set ${\bf P^-}$
of all polynomials in $\phi_\theta, \theta\in\lattice^-,$
are similarly defined.

We look into the $k$-th term in the right hand side of \eqr{JbyBB}.
Put
\[
 \psi = B^k\phi
\]
and write
\[
 (\phi,{B^*}^kB^k\phi)=||\psi||^2
  =\sum_{\theta_\Lambda,...,\theta_{k+1}}\psi_{\theta_\Lambda,...,\theta_{k+1}}^2 \ .
\]

Suppose that $l>k$.
Note that
\[
 \psi_{\theta_\Lambda,...,\theta_{k+1}}\in\left\{
   \begin{array}{ll}
    {\bf P^+},& \theta_l=1\ , \\
    {\bf P^-},& \theta_l=0\ .
   \end{array}\right.
\]
Putting
\alb
 \psi^2_{+}&=\mathop{\sum_{\theta_\Lambda,...,\theta_{k+1}=0,1}}_{\theta_l=1}
   \psi_{\theta_\Lambda,...,\theta_{k+1}}^2 
\ ,\nnb
 \psi^2_{-}&=\mathop{\sum_{\theta_\Lambda,...,\theta_{k+1}=0,1}}_{\theta_l=0}
   \psi_{\theta_\Lambda,...,\theta_{k+1}}^2 \ , \nn
\ale
we have  $\psi^2_-=\rho_l\psi^2_+$ and hence
\[
 ||\psi||^2={\psi^2_+}+{\psi^2_-}
   ={\psi^2_+}+\rho_l{\psi^2_+} \ .
\]
Then, 
$(\phi,{B^*}^kB^k\phi)=||\psi||^2$ makes a reflection-positive interaction.

Suppose that $l\le k$.
We can decompose $\psi$ as
\[
 \psi=\psi^+ + \psi^-\ ,\quad
  \psi^\pm\in{\bf P^\pm} \ ,
\]
with 
\[
\psi^-=\rho_l\psi^+\ .
\]
Then, we have
\[
 ||\psi||^2=||{\psi^+}+\rho_l{\psi^+}||^2
  =||{\psi^+}||^2+\rho_l||{\psi^+}||^2
   +2\sum_{\theta_\Lambda,...,\theta_{k+1}}
    \psi^+_{\theta_\Lambda,...,\theta_{k+1}}
    \rho_l\psi^+_{\theta_\Lambda,...,\theta_{k+1}} \ .
\]
This means that
$(\phi,{B^*}^k B^k \phi)=||\psi||^2$
makes a reflection-positive interaction.
\qed\prfe

\subsection{Hierarchical Laplacian}\appl{C}
We decompose $J$ as
\alb\eql{J=lap+mu}
 J&=\triangle_\Lambda+\mu_0I
\ale
so that 
\eqb\eql{lap1=0}
 \triangle_\Lambda {\bf 1}_\Lambda=0
\eqe
holds, where $\mu_0$ is a constant and
${\bf 1}_\Lambda={}^t(1,1,\cdots,1)\in\reals^{\lattice}$.
In fact, using $B{\bf 1}_{\Lambda-k}=\sqrt2{\bf 1}_{\Lambda-k-1}$, we have
\[
\mu_0
=\dfrac{1}{||{\bf 1}_{\Lambda}||^2}({\bf 1}_\Lambda,J{\bf 1}_\Lambda)
=\sum_{k=1}^{\Lambda}\dfrac{\beta_k}{\omega^k},
\]
and hence
\[
 -\triangle_\Lambda
=\sum_{k=1}^{\Lambda}\dfrac{\beta_k}{\omega^k}(-{B^*}^{k}B^k+I) 
=\sum_{k=1}^{\Lambda}\dfrac{\beta_k}{\omega^k}(-P_k+I) 
\ ,
\]
where
\[
 P_k={B^*}^kB^k\ ,\quad k=0,1,2,\cdots,\Lambda  \ .
\]
Since $BB^*=I$,
it holds that
\alb
 P_jP_k&=P_{\max(j,k)}\ ,\\
 P_j^*&=P_j \ ,
\ale
namely, $P_j$'s are orthogonal projections satisfying
\[
 P_\Lambda<P_{\Lambda-1}<\cdots<P_1<P_0=I\ .
\]
Define $Q_j, j=0,1,2,\cdots,\Lambda,$ by
\alb
 Q_j&=P_j-P_{j+1}\ ,\quad j=0,1,2,\cdots,\Lambda-1 \ , \nnb
 Q_\Lambda&=P_\Lambda \ .\nn
\ale
Then, $Q_0,Q_1,\cdots,Q_\Lambda$ constitute orthogonal projections with
\alb
 Q_jQ_k&=\delta_{jk}Q_j\ , \nnb
\sum_{j=0}^{\Lambda}Q_j&=I\ .\nn
\ale
In this notation, 
matrices $-\triangle_\Lambda+m^2I$ and $C=(-\triangle_\Lambda+m^2I)^{-1}$ 
are written as 
\alb\eql{-lap+m2}
 -\triangle_\Lambda+m^2I &= \sum_{k=0}^{\Lambda} y_kQ_k 
\ ,\quad \mbox{for } m^2\ge0
\ ,\\
\eql{Cexpression}
 C&= \sum_{k=0}^\Lambda \dfrac{1}{y_k}Q_k
\ ,\quad \mbox{for } m^2>0\ ,
\intertext{where}
\eql{yk}
 y_k&=\left\{\begin{array}{ll}
  \displaystyle\sum_{j=k+1}^\Lambda\dfrac{\beta_j}{\omega^j}+m^2\ ,
    &k=0,1,2,\cdots,\Lambda-1\ ,\\  &\\
  m^2\ ,&k=\Lambda\ .
 \end{array}\right.
\ale

\lemb\leml{choiceofm2}
\begin{enumerate}
\item
The matrix $-\triangle_\Lambda$ is positive semi-definite
and its eigenvalues lie in $[0,\mu_0]$.
\item For $l\ge1$, it holds that
\[
 0<\Tr C^l\le\dfrac{2^\Lambda}{m^{2l}} \ .
\]
\end{enumerate}
\leme
\prfb
(1) Put $m^2=0$ in \eqr{-lap+m2} and \eqr{yk}.
Then, we see that the smallest eigenvalue of $-\triangle_\Lambda$ is $0$
and the largest one is
\[
 \sum_{j=1}^\Lambda\dfrac{\beta_j}{\omega^j}=\mu_0 \ .
\]
(2) 
Note that \eqr{Cexpression} implies
\eqb\eql{TrC}
 \Tr C^l
= \sum_{k=0}^\Lambda \dfrac{1}{y^l_k}\Tr Q_k
=\sum_{k=0}^{\Lambda-1}\dfrac{2^{\Lambda-k-1}}{y_k^l}+\dfrac{1}{y_\Lambda^l}
 \ ,\quad
 l=1,2,\cdots \ .
\eqe
Since \eqr{yk} implies $y_k\ge m^2$,
we can estimate the right hand side of \eqr{TrC}.
\qed\prfe

Based on \lemr{RP} and \lemr{choiceofm2},
we can apply the $1/N$ expansion \cite{K} to the hierarchical system
and obtain \prpr{K}.

\prfofb{\lemr{C}}
(1) Note that \eqr{Cexpression} and the equality
\[
 B^\Lambda Q_j{B^*}^\Lambda=\left\{
 \begin{array}{ll}
  0\  ,& 0\le j\le \Lambda-1\ ,\\
  1\  ,& j=\Lambda\ ,
 \end{array}\right. 
\]
yield
\[
 B^\Lambda C{B^*}^\Lambda=\dfrac{1}{m^2} \ .
\]
This means \eqr{csus}.

\noindent
(2) Since
\[
 (Q_j)_{\theta\theta}=\left\{
 \begin{array}{ll}
  2^{-j-1},& 0\le j\le \Lambda-1\ ,\\
  2^{-\Lambda},&j=\Lambda\ ,
 \end{array}\right. 
 \quad \theta\in\lattice\ ,
\]
\eqr{Cexpression} implies
\[
 C_{\theta\theta}
 =\sum_{k=0}^{\Lambda-1}\dfrac{1}{2^{k+1}y_k}+\dfrac{1}{2^{\Lambda}y_\Lambda}
 \ ,\quad
 \theta\in\lattice\ .
\]
Now, under \eqr{betak} we have 
\[
 y_k=\left\{\begin{array}{ll}
  \dfrac{1}{2\omega^\Lambda}(\omega^{\Lambda-k}-\omega+4t+2\omega^{\Lambda}m^2)\ ,
    & 0\le k\le \Lambda-1 \ , \\
  &\\
  m^2\ , & k=\Lambda \ .
  \end{array}\right.
\]
Then, it holds that
\[
 C_{\theta\theta}
=(\dfrac{\omega}{2})^\Lambda\sum_{j=1}^\Lambda
  \dfrac{2^j}{\omega^j-\omega+4t+2\omega^\Lambda m^2}+
  \dfrac{1}{2^\Lambda m^2} \ .
\]
\qed\prfofe


\end{document}